\documentclass[fleqn,usenatbib]{mnras}

\usepackage{newtxtext,newtxmath}
\usepackage{units}
\usepackage[T1]{fontenc}
\usepackage[dvipsnames]{xcolor}

\DeclareRobustCommand{\VAN}[3]{#2}
\let\VANthebibliography\thebibliography
\def\thebibliography{\DeclareRobustCommand{\VAN}[3]{##3}\VANthebibliography}

\newcommand{\ap}[1]{'#1}

\usepackage{graphicx}	
\usepackage{amsmath}	

\title[Unequal Mass Inspirals]{The dynamics and electromagnetic signatures of accretion in unequal mass binary black hole inspirals}

\author[M. Clyburn and J. Zrake]{
Madeline Clyburn$^{1}$\thanks{E-mail: clyburn@clemson.edu} \&
Jonathan Zrake$^{1}$\thanks{E-mail: jzrake@clemson.edu} \\
$^{1}$Department of Physics and Astronomy, Clemson University, Clemson, SC 29634, USA\\
}

\date{Accepted XXX. Received YYY; in original form ZZZ}

\pubyear{2024}

\begin{document}
\label{firstpage}
\pagerange{\pageref{firstpage}--\pageref{lastpage}}
\maketitle

\begin{abstract}
We present a theoretical study of the gravitational wave (GW) driven inspirals of accreting black hole binaries with mass $M = 10^7 M_\odot$ and mass ratios between $10^{-3}$ and $10^{-1}$. Our results are based on analytic estimates, and grid-based hydrodynamics simulations run for many thousands of binary orbits before the merger. We show that the GW inspiral is evident in the light curves and color evolution of a binary-hosting quasar over years to decades before a merger. The long-term electromagnetic (EM) signature is characterized by a gradual UV brightening and X-ray dimming, followed by an X-ray disappearance hours to days before the GW burst, and finally, a years-like re-brightening as the disk relaxes and refuels the remnant black hole. These timescales are surprisingly insensitive to the normalization of the kinematic viscosity in the disk. The spectrum of quasi-thermal disk emission shows two peaks: one in the UV and another in the X-ray, associated with the outer and circum-secondary disks, respectively; emission from the inner disk is suppressed because the secondary consumes most of the inflowing gas. We discuss implications for real-time and archival EM follow-up of GW bursts detected by \emph{LISA}.
\end{abstract}


\section{Introduction}
The gravitational wave (GW) driven inspirals of massive black hole binaries (MBHB\ap{s}) are high-priority targets for the future space-based GW detectors such as \textit{LISA}, \textit{Taiji}, and \textit{TianQin} \citep{AmaroSeoane2023, Ruan2018, Gong2021}. Within the next fifteen years, \textit{LISA} will observe the GW\ap{s} emitted by MBHB\ap{s} in the final hours to months of their lives \cite[e.g.][]{Sesana2021}, including the GW burst produced as they merge, and the ring-down of the final black hole remnant \citep{Baibhav2020, Pitte2023}. These observations will directly constrain the binary component masses and spins \citep[e.g.][]{Trias2008}, and could reveal the presence of a circumbinary gas disk \citep{Yunes2011, Derdzinski2019, Derdzinski2021, Garg2024}.

The science potential of a \emph{LISA} detection is greatly increased if the host galaxy can be identified. The host morphology could reveal evidence of recent galaxy mergers \cite[e.g.][]{Hughes2002, Villalba2022}, potentially constraining the timescale for MBHB formation \citep{Begelman1980}. Also, the host redshift, combined with the GW luminosity distance, yields a measurement of the Hubble constant \citep{Schutz1986}. However, \emph{LISA} observations of MBHB inspirals could be localized to a region of $1-10 \deg^2$, generally too large to identify the host galaxies of individual events prior to merger \citep{Mangiagli2020}. It should be noted that post-merger, the \textit{LISA} localization region can be potentially as small as $0.01 \deg^2$. Nevertheless, host galaxy identification in the pre-merger phase will almost certainly require an EM counterpart, and pre-merger galaxy selection can uniquely allow for multi-wavelength monitoring of the source during and after the merger.

The MBHB\ap{s} that will merge as \textit{LISA} sources likely exist in the gas-rich nuclear regions of post-merger galaxies \citep{barnes1996} and could thus be surrounded by an optically thick accretion flow. The flow would consist of an outer (circumbinary) disk, an inner disk around the primary (more massive) black hole and a smaller ``minidisk'' around the secondary (less massive) black hole. The quasi-thermal radiation released from the accretion flow could exhibit distinctive temporal or spectral characteristics, which may aid in the identification of a host galaxy, either in archival data from time-domain EM surveys or with contemporaneous tiling of the \emph{LISA} error box.

Predicting such EM signatures requires a detailed understanding of how the accretion flow evolves in time throughout the binary inspiral, and many studies have now been carried out based on (magneto-)hydrodynamics simulations in both 2D \citep[e.g.][]{Noble2012, Farris2015, Tang2018, Derdzinski2019, Derdzinski2021, Dittmann2023, Krauth2023b} and 3D \citep[e.g.][]{Farris2010, Farris2011, Farris2012,  Bowen2018, Bowen2019, Pereira2019, Combi2021, Gutierrez2022, Ruiz2023, Avara2023, Cocchiararo2024, Franchini2024}. Each of these is based on equal-mass binary inspirals, except for \cite{Ruiz2023}, which includes a model with a mass ratio as low as $q \equiv M_2/M_1 = 0.25$, and \cite{Pereira2019} and \cite{Derdzinski2019, Derdzinski2021}, which include intermediate mass-ratio inspirals \citep[so-called IMRI\ap{s};][]{Arca-Sedda2020}, in the range of $q = 10^{-2}$ to $10^{-4}$. The study led by Pereira reports the time-evolving mass accretion rates, but not EM signatures, and the studies led by Derdzinski analyze the magnitude of gas-induced deviations from the vacuum post-Newtonian inspiral, but also do not characterize the EM signatures. Very recently, \cite{Cocchiararo2024} reported EM signatures for mass ratios as small as $q=0.1$.

In this paper, we use multi-dimensional hydrodynamics simulations to characterize the EM signatures of unequal mass inspirals, with $q$ in the range of $10^{-3}$ to $10^{-1}$. Our interest in unequal mass inspirals is motivated by the possibility that a large fraction of the MBHB population could have significantly unequal masses \citep{Sesana2012, Bellovary2019}, and that accretion in low-mass ratio binaries seems to operate with reduced stochastic variability, especially below $q \sim 0.01- 0.04$ \citep[e.g.][]{Farris2014, Shi2015, DOrazio2016, Duffell2020, Dittmann2024}, which could allow for periodic or secular variability signatures to stand out more clearly.
IMRI\ap{s} are also of particular interest because the lower-mass component black holes will be in the range of $10^{3-5} \ \rm{M}_\odot$, meaning that \emph{LISA} observations of them would directly probe a population of intermediate-mass black holes, which has so far been elusive \citep[e.g.][]{Miller2009}.

Accreting IMRI\ap{s} might also present distinctive spectral signatures to distinguish them from MBHB\ap{s} with nearly equal component masses. This is because the less massive black hole is generally expected to consume a majority of the inflowing gas as seen in numerical simulations with thin disks \citep[e.g.][]{Lubow1999, Duffell2020, Munoz2020, Siwek2023a}, an effect referred to in literature as preferential accretion \citep{Munoz2016}. The effect implies that emission from the secondary minidisk could dominate the system's overall EM luminosity or that the spectral energy distribution of quasi-thermal disk emission could exhibit multiple peaks. Also, if the degree of preferential accretion changes throughout the late inspiral, that might imprint a distinctive color evolution that could be used to photometrically select host galaxy candidates from time-domain surveys. Thus, the main goal of this study is to measure the evolution of the component accretion rates $\dot M_1$ and $\dot M_2$ throughout a GW-driven inspiral, and use those measurements to approximate the color evolution of accreting IMRI\ap{s}.

To the best of our knowledge, the present study is the first to use multi-dimensional hydrodynamics simulations to model the EM emission of accreting IMRI\ap{s}. However, the earliest papers on EM emission from MBHB inspirals, which were based on one-dimensional disk models, focused specifically on systems with mass ratios $q \lesssim 0.1$. \cite{Armitage2002} used the ``standard disk'' equations \cite[e.g.][]{Shakura1973, Pringle1981} to model how the inner and outer disks evolve throughout the inspiral. They used an empirical term derived from the theory of disk-satellite interactions \citep{Goldreich1980, Lin1986} to model the gravitational torque applied to the disk by the secondary. Such treatments can lead to the formation of a vacuum gap surrounding the secondary's orbit, prohibiting mass from moving between the inner and outer disks or onto the secondary black hole. As the GW inspiral speeds up, the vacuum gap implies ``tidal squeezing'' of the inner disk toward the primary black hole. \cite{Armitage2002} speculated that this process (now sometimes referred to as a ``snowplow'' mechanism) force-feeds the larger black hole and could drive a massive outflow and EM flare in the very late inspiral. The studies of \cite{Lodato2009} and \cite{Chang2010} are similarly based on time-dependent one-dimensional disk models and also support the possibility of a snowplow mechanism.

However, the vacuum gap around the secondary's orbit is now understood to be an artifact of one-dimensional disk treatments. Two-dimensional calculations show that gas crosses the secondary orbit on so-called ``horseshoe'' orbits \citep{Lubow1999, Duffell2014, Fung2014}. It means that the inner disk gas could be expelled to the outer disk or consumed by the secondary and that a late-inspiral enhancement of the accretion rate may not take place \citep{Baruteau2012}. In fact, the opposite effect has been seen in simulations of equal-mass binaries \citep[e.g.][]{Farris2015, Dittmann2023, Krauth2023b, Franchini2024}; the binary accretion rate drops throughout the late inspiral. We refer to such mass starvation as ``viscous decoupling,'' an effect first described in \cite{Liu2003}. It happens because the binary eventually contracts faster than the inward viscous spreading of the outer disk, and this leads to a diminishing rate of mass supply to the binary components. It was also pointed out in \cite{Milosavljevic2005} that viscous decoupling of the binary should lead to a post-merger re-brightening as the disk spreads inwards following the merger and eventually begins fueling the black hole remnant. The re-fueling has been simulated for the case of post-merger equal-mass binaries in \cite{Farris2015} and \cite{Krauth2023b}, and we examine it in this study for the case of intermediate-mass ratios $q < 0.1$.

A cautionary note for readers of this paper and similar works; previous works have not always made the distinction between two different timescales in the binary-disk interaction. One timescale, the viscous decoupling timescale, as mentioned in the preceding paragraph, describes the time at which the binary merger rate begins to contract faster than the viscous spreading at the inner edge of the outer disk. Previous works \citep[e.g.][]{Sesana2012} have referred to this timescale as the ``disk freezing timescale.'' The second timescale is the time at which the binary switches from contraction/expansion due to the exchange of angular momentum with the gas disk to contraction due to GW emission. In this paper, we will refer to this timescale as the ``transition timescale,'' but others have referred to this timescale as a ``decoupling timescale'' \citep[e.g.][]{Armitage2002, Sesana2012}. These timescales are distinct and should be calculated separately; see Secs. \ref{subsec:dec} and \ref{subsec:trans} for derivations of the viscous decoupling timescale and transition timescale, respectively.

Our study is based on time-dependent numerical solutions of the two-dimensional (vertically averaged) Navier-Stokes equations, where gas is subject to the time-dependent gravitational potential of an unequal-mass black hole binary, undergoing a GW-driven post-Newtonian inspiral \citep{Peters1964}. We adopt a simplified thermodynamic treatment, in which the speed of sound is 
set to a small fraction, $1/\mathcal{M}$, of the nominal orbital speed ($\mathcal{M}$ is a global Mach number, selected to be in the range of $10-20$).
The bolometric EM luminosity is reasonably well predicted by measurements of the mass accretion rate, provided the flow is radiatively efficient. Owing to the use of simplified thermodynamic treatment, our simulations do not yield a self-consistent disk photosphere temperature. To model the color evolution, we have thus developed a ``three-disk'' toy model for the disk surface temperature, based on approximating the inner, outer, and secondary disks as stationary $\alpha$-disks \citep{Shakura1973}, with simulation-calibrated mass inflow rates. We believe this approach yields a reasonable first approximation of the color evolution of accreting IMRI\ap{s}. Models with self-consistent thermodynamics \cite[e.g.][]{WesternacherSchneider2022, WesternacherSchneider2023} for unequal mass binaries will be reported in a follow-up study.

Our paper is organized as follows. In Sec. \ref{sec:picture}, we describe the physical picture of the MBHB inspiral and provide analytical estimates for the EM emission from these systems. In Sec. \ref{sec:numerical}, we describe the equations of motion, initial conditions, and the numerical setup. Sec. \ref{sec:results} contains the key dynamical measurements obtained from the numerical simulations. Sec. \ref{sec:observables} includes the predictions of spectral evolution and light curves based on the three-disk spectral toy model and discusses prospects for EM detections of \textit{LISA} sources. Finally, in Sec. \ref{sec:discussion}, we summarize our results, discuss some caveats, and describe plans for future work.

\section{Accreting Unequal Mass Black Hole Binaries}
\label{sec:picture}

\subsection{Fiducial System}
\label{subsec:fiducial}
We adopt a fiducial black hole binary that will be detectable by \textit{LISA} out to a high redshift. The total mass of the fiducial system is $M = 10^7 \ \rm{M}_\odot$, the mass ratio is $q = 0.01$, and the orbit is circular. The assumption of a circular orbit is justified on the basis that in the late inspiral stage, the GW radiative back-reaction exceeds the gravitational forces of the gas, so the orbital eccentricity, driven up at wider binary separations \citep{Zrake2021, DOrazio2021, Siwek2023b}, has already been dissipated. At the leading post-Newtonian order, the rate of orbital contraction due to GW\ap{s} is given in \cite{Peters1964} as
\begin{equation}
    \dot{a}_{\rm gw} = -\frac{64 G^3M^3\eta}{5 c^5 a^3} \, ,
    \label{eq:adot}
\end{equation}
where $\eta \equiv q / (1 + q)^2 = M_1 M_2 / M^2$ and $a$ is the size of the orbit. If the fiducial binary were on track to merge in the next $50$ years, then it would presently have a semi-major axis $a_0 \simeq \unit[6.3]{AU}$ and an orbital period $P_0 \simeq \unit[1.8]{days}$. The time-to-coalescence is denoted as $t_{\rm minus} \equiv \tau / 4$, where $\tau \equiv - a / \dot a_{\rm gw}$ is the nominal orbital contraction timescale. The fiducial system is very mildly relativistic, with the orbital velocity roughly equivalent to $0.17c$ and gravitational radius $r_g \equiv GM / c^2$ about $64$ times smaller than the initial semi-major axis. For later reference, the three (small) relativistic parameters $\varv/c$, $r_g/a$, and $P/\tau$ are mutually related via Eqn. \ref{eq:adot},
\begin{equation}
    \left(\frac{\varv}{c}\right)^2 = \frac{r_g}{a} = \left(\frac{5P}{128 \pi \eta \tau}\right)^{2/5} \, .
    \label{eq:notm-rg-v}
\end{equation}
The Eddington luminosity of a $10^7 M_\odot$ black hole is $L_{\rm edd} \simeq 1.15 \times \unit[10^{45}]{erg/s}$, corresponding to a bolometric absolute magnitude of $M_{\rm bol} \simeq -24$. For reference, such modest-luminosity AGN will be detectable by \textit{LSST} out to redshifts of $z \simeq 5.5$. \citep{LSSTbook}.  The \textit{LISA} detection horizon for these events will be at redshifts of $z \simeq 20$ \citep{Sesana2021}.

\subsection{Physical Picture and Spectral Appearance}
\label{subsec:spectrum}
The binary is surrounded by a thin, centrifugally supported, optically thick gas flow. In binaries with a small mass ratio, the overall flow is similar to that of planet-forming disks. It comprises an outer circumbinary disk, a low-density annular gap surrounding the secondary's orbit, and an inner disk around the primary.

The accretion flow appears as a multi-temperature blackbody; the spectrum is determined by the distribution of the local disk photosphere temperatures. To get a sense for the composite spectrum of the accretion flow, including the outer disk, the inner disk, and the secondary minidisk, it is instructive to consider each of these disks as a steady state $\alpha$-disk \citep{Shakura1973}, and examine the composite quasi-thermal emission spectrum. The characteristic temperatures are given by
\begin{eqnarray}
    T_{\rm outer} &=& (3 G M \dot M / 8 \pi a^3 \sigma)^{1/4} \nonumber \\
    T_i &=& (3 G M_i \dot M_i / 8 \pi R_i^3 \sigma)^{1/4} \propto \dot{M}_i^{1/4} M_i^{-1/2}
    \label{eq:temp}
\end{eqnarray}
where $i=1$ for the primary and $i=2$ for the secondary, and $R_i = 6 G M_i / c^2$ are the inner cutoff radii of the component disks. For the fiducial system discussed in Sec. \ref{subsec:fiducial}, accreting at the Eddington limit of the primary\footnote{If the circumbinary disk carries mass inwards at a large fraction of the primary's Eddington limit, the secondary would need to accrete in a super-Eddington regime to consume a significant fraction of the overall mass flow. Thus, our assumption that the secondary disk remains thin is not technically compatible with assumptions elsewhere that the binary accretion rate is comparable to the Eddington rate. This very interesting aspect of accretion in unequal-mass binary systems needs to be addressed but is out of the scope of the present study. Most AGN accrete below Eddington, so we may be overestimating typical luminosities here while not violating the physical assumption of a thin secondary disk.}
, with $\dot M_2 / \dot M_1 = 10$, the characteristic temperatures of the three disks are
\begin{eqnarray*}
    kT_{\rm outer} &\simeq& \unit[28]{eV} \\
    kT_{1} &\simeq& \unit[41]{eV} \\
    kT_{2} &\simeq& \unit[1.3]{keV} \, .
\end{eqnarray*}
The secondary black hole has a more compact disk and typically accretes faster, so the secondary disk emission is bluer and brighter than the primary and outer disks.


%
\begin{figure}
    \centering
    \includegraphics{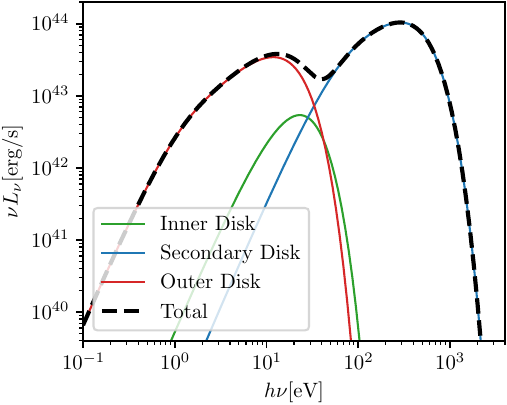}
    \caption{Toy model of the spectral energy distribution of thermal emission produced by an accretion flow around a $M=10^7 M_\odot$ binary with mass ratio $q=0.01$. The model assumes that the three components of the accretion flow: inner disk (green), secondary disk (blue), and outer disk (red) can each be characterized as a steady state $\alpha$-disk, with radial temperature profile consistent with Eqn. \ref{eq:temp}.
    }
    \label{fig:binary_spectrum}
\end{figure}

Fig. \ref{fig:binary_spectrum} shows a toy model spectrum based on this picture and the temperature profiles in Eqn. \ref{eq:temp}. The inner disk extends to $r=a$, and the secondary disk extends to the Hill radius $r_{\rm hill} \simeq a (q/3)^{1/3}$. The outer disk starts at $r=a$ and extends to infinity. The preferential accretion parameter is set to $\dot M_2 / \dot M_1 = 10$. The spectrum is double-peaked, with the UV and X-ray bumps formed by the outer and secondary disks, respectively. The inner disk is out-shined by the other components, a consequence of the secondary black hole dominating the accretion. Note that a double-peaked spectrum similar to that shown in Fig. \ref{fig:binary_spectrum} is also presented in \cite{Chang2010}, but the reason is different. In that study, the secondary does not accrete, and the two spectral bumps are produced by the inner and outer disks.

\subsection{Viscous Decoupling Timescale}
\label{subsec:dec}
At some point during the GW inspiral, the GW contraction rate $\tau^{-1}$ exceeds the viscous relaxation rate \citep[e.g.][]{Liu2003, Milosavljevic2005}. After that time, the disk is said to have ``viscously decoupled'' from the binary, as it can no longer spread viscously inwards as fast as the speed $\dot a_{\rm gw}$ of the binary contraction. The secular evolution of the brightness and color of the accretion flow around the binary could be significantly altered by viscous decoupling. Note, that previous works \cite[e.g.][]{Sesana2012} have referred to this timescale as the ``disk freezing timescale.'' Also, the term ``decoupling timescale'' has been used in other works to refer to the time after which the GW power exceeds the rate of energy loss due to tidal interactions with the disk. In this paper, we refer to the transition from gas to GW driving as the ``transition timescale.'' The transition timescale is estimated later in Sec. \ref{subsec:trans}.

The nominal viscous relaxation timescale of the accretion disk is given by
\begin{equation}
    t_{\rm visc}(r) = \frac{2}{3}\frac{r^2}{\nu(r)} \, ,
    \label{eq:tvisc}
\end{equation}
where $\nu(r)$ is the kinematic viscosity at radius $r$. In a standard $\alpha$-disk with constant aspect ratio $h/r$, the viscosity coefficient can be written as $\nu(r) = \bar \nu \sqrt{G M r}$. Here $\bar \nu \equiv \alpha / \mathcal{M}^2$, where $\alpha \lesssim 0.1$ represents the Reynolds stress associated with small-scale (numerically unresolved) turbulence in the disk \citep{Shakura1973}, and $\mathcal{M}$ is the orbital Mach number. In addition to the fiducial binary parameters introduced in Sec. \ref{subsec:fiducial}, we also define a set of fiducial disk parameters $\alpha = 0.1$ and $\mathcal{M} = 10$.

\begin{figure*}
  \begin{center}
  \includegraphics{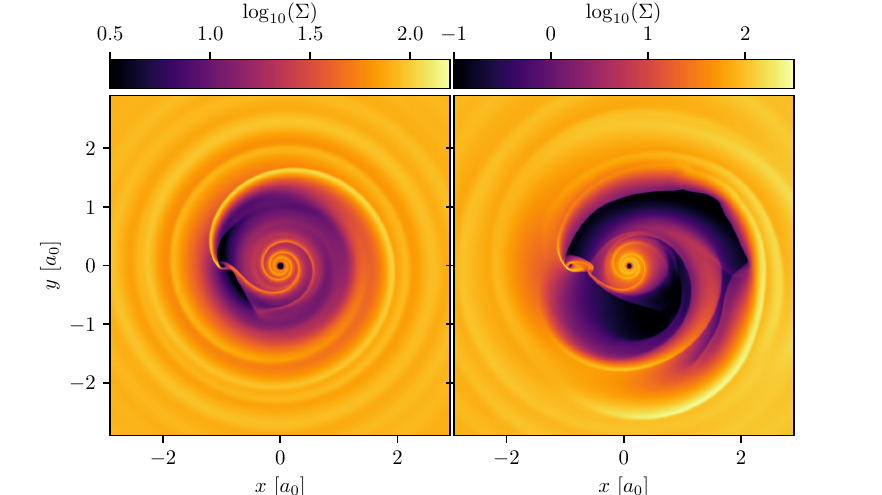}
  \end{center}
  \caption{Image of the surface density of the disk around a MBHB with mass ratio $q=0.01$ (left) and $q=0.1$ (right), accreting from a circumbinary gas disk. The image is from a simulation snapshot taken well after the disk is viscously relaxed but before the binary separation has been reduced much from $a_0$. The axes show the $x-y$ plane in units of $a_0$.}
  \label{fig:sailfish}
\end{figure*}

The ratio of the viscous time near the binary to the GW inspiral time is
\begin{equation}
    \frac{t_{\rm visc}(a)}{\tau} = \frac{128 G^3M^3\eta}{15\nu(a)c^5a^2} \propto a^{-5/2} \, .
    \label{eq:tvisc-tgw}
\end{equation}
Note that we have chosen to equate the inspiral timescale to the viscous timescale at $r = a$ for the viscous decoupling condition. This choice seems more appropriate when the mass ratio is small (see Figs. \ref{fig:sailfish} and \ref{fig:SigmaProfile}) rather than at $r = 2.5a$, which is typically used as the radius for the outer disk cavity edge in equal-mass binary studies.
This ratio, as shown in Eqn. \ref{eq:tvisc-tgw}, starts off small and increases as the inspiral proceeds. It exceeds unity at the viscous decoupling radius $a_{\rm dec}$,
\begin{eqnarray}
\label{eq:adec}
    a_{\rm dec} &=& r_g\bigg(\frac{128 \eta}{15 \bar{\nu}} \bigg)^{2/5} \\
                &\simeq& \unit[0.6]{AU} \times
    \bigg(\frac{M}{10^7 M_{\odot}}\bigg)
    \bigg(\frac{\eta}{10^{-2}}\bigg)^{2/5}
    \bigg(\frac{\bar{\nu}}{10^{-3}}\bigg)^{-2/5} \, , \nonumber
\end{eqnarray}
which we have written here, first in terms of $r_g$, and then for a typical \textit{LISA} binary with the fiducial parameters described in Sec. \ref{subsec:fiducial}. 
The method we have used to determine the decoupling separation in Eqn. \ref{eq:adec} has been questioned in works such as \cite{Dittmann2023}, where the authors show that a more accurate method to derive the decoupling separation is to equate the binary contraction speed to the viscous drift speed rather than equating timescales. Nevertheless, our result in Eqn. \ref{eq:adec}, is the same regardless of equating timescales or speeds. This is due to our assumption that $t_{\rm visc} = \tau_{\rm gw}$ while \cite{Dittmann2023} assumes $t_{\rm visc} = \tau_{\rm gw} / 4$. Thus the viscous decoupling scales we derived here are equivalent to those in Eqns. 5 and 6 of \cite{Dittmann2023}, for their case of $\xi=1$ and $\lambda=2$.

The residence time for the fiducial system at the viscous decoupling stage is
\begin{eqnarray}
    \label{eq:t_dec}
    \tau_{\rm dec} &=& \frac{5}{64}\frac{r_g}{c\eta}\bigg(\frac{128 \eta}{15 \bar{\nu}} \bigg)^{8/5} \\
    &\simeq& \unit[5.4]{days} \times
    \bigg(\frac{M}{10^7 M_{\odot}}\bigg)
    \bigg(\frac{\eta}{10^{-2}}\bigg)^{3/5}
    \bigg(\frac{\bar{\nu}}{10^{-3}}\bigg)^{-8/5} \, , \nonumber
\end{eqnarray}
which corresponds to a time-to-coalescence of $t_{\rm minus} \simeq 1.4 \ \text{days}$. The orbital period at the viscous decoupling radius is
\begin{equation}
    P_{\rm dec} = \unit[1.2]{hours} \times
    \bigg(\frac{M}{10^7 M_{\odot}}\bigg)
    \bigg(\frac{\eta}{10^{-2}}\bigg)^{3/5}
    \bigg(\frac{\bar{\nu}}{10^{-3}}\bigg)^{-3/5} \, .
    \label{eq:Pdec-ex}
\end{equation}
Eqn. \ref{eq:t_dec} shows the timescale at which the binary is generally expected to become starved of mass supply. Note the apparent high sensitivity to the viscosity suggests that thinner $\alpha$-disks might viscously decouple much earlier than warmer disks; if $\bar{\nu}$ were instead $10^{-5}$ ($\mathcal{M} = 100$), Eqn. \ref{eq:t_dec} indicates the viscous decoupling timescale would then be on the order of years rather than days. In Sec. \ref{subsec:mdots}, we show that this expectation is not supported by our simulations; we find instead that viscous decoupling extends over thousands of $\tau_{\rm dec}$, implying that colder and warmer disks alike could show signs of the inspiral long before the nominal decoupling time.

Measurements of the preferential accretion parameter $\dot M_2 / \dot M_1$ are reported in a number of studies \citep[e.g.][]{Lubow1999, Young2015, Munoz2020, Duffell2020, Siwek2023a}, and these generally find values of $\dot M_2 / \dot M_1$ in the range of $1 - 10$, when the mass ratio is larger than roughly $10^{-3}$. All of those studies were based on simulations with a viscously relaxed disk. In Sec. \ref{subsec:mdots}, we present calculations revealing how the preferential accretion parameter evolves throughout the long-term viscous decoupling.

\subsection{An Argument Against Tidal Squeezing}
\label{subsec:snowplow}
As the disk loses its viscous coupling to the binary, the feeding rates $\dot M_1$ and $\dot M_2$ evolve away from their steady-state values. These long-term trends will largely determine the IMRI brightness and color evolution before coalescence. However, the feeding rates are not easy to predict without detailed simulations. The main question here is whether the GW burst is preceded by a period of AGN brightening or dimming. \cite{Armitage2002} computed the evolution of a one-dimensional Keplerian disk with an embedded low-mass secondary black hole ($q=0.01$) on a GW-driven inspiral. They accounted for the gravitational torque of the secondary on the inner and outer disks, using a prescription for Lindblad torques due to \cite{Lin1986}. That prescription leads to the formation of an evacuated gap around the secondary orbit, which fully insulates the mass flow between the inner and outer disks and prohibits any movement of gas onto the secondary black hole. The result, which can be seen, for example, in Fig. 3 in \cite{Armitage2002}, is that the inner disk is squeezed toward the primary as the secondary spirals in.

Tidal squeezing would require the secondary to exert sufficient gravitational torque to the inner disk to lower the orbits of gas parcels faster than $\dot a_{\rm gw}$. It implies the positive torque, $\dot J_{\rm in}$, exerted by the inner disk on the secondary, must exceed
\begin{equation}
    \dot J_{\rm plow} \equiv \pi a^2 \Sigma(a) \times \Omega(a) a^2 / \tau_{\rm gw}
    \simeq \frac{64 \eta \dot M r_g^3 c}{15 \bar \nu a^2} \, ,
    \label{eq:Jdotplow}
\end{equation}
where the second equality is precise in a steady state with $\dot M = 3 \pi \nu(r) \Sigma(r)$. The dependence of this torque on $a^{-2}$ indicates that to squeeze the inner disk would require the torque to diverge leading up to merger. The gravitational torque coming from the inner disk can be written as $\dot J_{\rm in} = \ell_{\rm in} J_2 \dot M / M$ where $J_2 \simeq M_2 \Omega a^2$ is the secondary angular momentum, and $\ell_{\rm in}$ is a dimensionless ``eigenvalue'', which is typically not greater in magnitude than about $10$ which is confirmed for binary mass ratio as low as $q=0.01$ \citep[e.g.][]{Duffell2020}. It is then straightforward to see that $\dot J_{\rm plow}$ (the torque required to squeeze the inner disk) exceeds $\dot J_2$ some $1 / (6 \pi \eta \bar \nu \ell_{\rm in})$ orbits before merger. In the fiducial binary and disk system with $q \approx \eta \simeq 0.01$, $\bar{\nu} = 10^{-3}$, and torque eigenvalue of order unity, this corresponds to $\sim\!5000$ orbits or $\sim\!25$ years before merger. Thus, on dynamical grounds, we expect that in the final thousands of orbits, the inner disk gas is either funneled to the outer disk or falls onto the secondary. It would require an unrealistically large gravitational torque to confine the inner disk inside the secondary's orbit in the late inspiral. In Sec. \ref{subsec:jdot}, we present numerical calculations confirming these estimates.

\subsection{Transition From Gas to GW Driving}
\label{subsec:trans}
For much of the binary lifetime, the orbital contraction is thought to be governed by coupling to the circumbinary disk \citep[e.g.]{Begelman1980}. As the binary contracts and the GW power increases, there is a transition from gas to GW driving \citep{Escala2005, Lodato2009, Cuadra2009, Munoz2020, Bortolas2021}. Once the binary reaches that separation, which we denote here as $a_{\rm trans}$, an inspiral towards merger is inevitable as GW\ap{s} remove orbital energy from the binary at a runaway rate. Note that some authors use different terminology for this separation and timescale, for example \cite{Sesana2012} simply calls this ``decoupling.'' We adopted the term ``transition separation'' for the switch to GW driving, in order to reserve the term decoupling for the process of viscous decoupling.

The transition separation occurs where the rate of orbital contraction by GW radiation (Eqn. \ref{eq:adot}) is equivalent to the rate of orbital contraction by gas i.e. $\dot{a}_{\rm gas} = \dot{a}_{\rm gw}$. Gas-induced changes of the semi-major axis are prescribed here in terms of another dimensionless constant $\epsilon$, as $\dot{a}_{\rm gas} = -\epsilon a \dot M / M$. Equating $\dot{a}_{\rm gw}$ and $\dot{a}_{\rm gas}$, we find the separation at which GW\ap{s} begin to dominate the inspiral,
\begin{eqnarray}
    \label{eq:atrans}
    a_{\rm trans} &=& r_g \bigg(\frac{64 \eta c^3}{5 G \dot{M}\epsilon}\bigg)^{1/4} \\
                &\simeq& \unit[94]{AU} \times
    \bigg(\frac{M}{10^7 M_{\odot}}\bigg)^{3/4}
    \bigg(\frac{\eta}{10^{-2}}\bigg)^{1/4}
    \bigg(\frac{\epsilon}{5.0}\bigg)^{-1/4} \dot{m}^{-1/4} \, , \nonumber
\end{eqnarray}
where $\dot m \equiv \dot M / \dot M_{\rm edd}$ is the mass accretion rate normalized by the Eddington accretion rate and $\epsilon \approx 5.0$ is consistent with results from simulations \citep[e.g.][]{Tiede2020, Dittmann2024, Tiede2024c}. The timescale for the transition from gas to GW-driven inspiral is $\tau_{\rm trans} \equiv \tau (a_{\rm trans}) = t_{\rm sal} / \epsilon \dot m$, where $t_{\rm sal} = \unit[5 \times 10^7]{years}$ is the Salpeter time. This nominal timescale for the end of the gas-driven evolutionary phase holds when the orbital decay follows $\dot a / a \sim -\dot M / M$ and the mass accretion rate is a fraction of the Eddington accretion rate.

Therefore, \emph{LISA} inspirals will be generally deep in the GW-driven regime, and it justifies the choice of circular binary orbits in this study. Even though a detectable level of orbital eccentricity could be left over from a gas-driven evolutionary phase in \emph{LISA} sources \citep{Armitage2005, Roedig2011, Zrake2021, Siwek2023b, Garg2024, Valli2024}, from the point of view of the circumbinary disk interaction years before a merger, the orbit can be considered as nearly circular.

\subsection{Post-Merger Viscous Closing}
\label{subsec:viscous-closing}
Viscous decoupling implies the binary merges in a relatively low-density ``hole'' in the disk. Here we estimate the timescale for the hole to be refilled following the merger, noting that such effect has been surmised to lead to a late-time brightening, or ``merger afterglow'' \cite[e.g.][]{Milosavljevic2005}. The afterglow is predicted to begin a time span $t_{\rm refill}$ following the merger, given by the viscous relaxation timescale at the radius where the binary became viscously decoupled,
\begin{eqnarray}
\label{eq:t_refill}
    t_{\rm refill} &\equiv& \frac{3}{4} \ t_{\rm visc}(a_{\rm dec}) = \frac{1}{2}\frac{r_g}{c\bar{\nu}}\bigg(\frac{128 \eta}{15 \bar{\nu}} \bigg)^{3/5} \\
    &\simeq& \unit[4.1]{days} \times
    \bigg(\frac{M}{10^7 M_{\odot}}\bigg)
    \bigg(\frac{\eta}{10^{-2}}\bigg)^{3/5}
    \bigg(\frac{\bar{\nu}}{10^{-3}}\bigg)^{-8/5} \, .  \nonumber
\end{eqnarray}
This estimate is analogous to one given in \cite{Milosavljevic2005}. Note that it suggests the remnant black hole begins refueling on a timescale that is quite sensitive to the disk viscosity; in an $\alpha$-viscosity model, the refueling timescale would appear to increase from days for $\mathcal{M}=10$ to decades for $\mathcal{M}=100$. We show in Sec. \ref{subsec:remnant} that such sensitivity is not supported by the simulations, instead the remnant begins fueling over year-like timescales, with surprisingly little sensitivity to the normalization of the kinematic viscosity.

\subsection{Importance of Self-Gravity}
\label{subsec:assumptions}
Our simulations do not include the self-gravity of the disk gas, so it is important to check when that assumption could be valid. The importance of self-gravity can be estimated from the
Toomre parameter, $Q = c_s \Omega / \pi G \Sigma$ for a Keplerian disk; disks are generally unstable to fragmentation instabilities if $Q \lesssim 1$. When $\alpha=0.1$ and $\mathcal{M} = 10$ the Toomre parameter is found to be
\begin{equation}
    Q = 6 \times 10^5 \ \bigg(\frac{M}{10^7 M_{\odot}}\bigg)^{1/2}
    \bigg(\frac{\bar{\nu}}{10^{-3}}\bigg)
    \bigg(\frac{r}{10a}\bigg)^{-3/2} \dot{m} \, .
    \label{eqn:Toomre}
\end{equation}
Colder disks are more unstable to gravitational fragmentation, however Eqn. \ref{eqn:Toomre} indicates that even when the Mach number is on the order of hundreds, the disk still has $Q \gg 1$ out to $r \sim 10a$. The effect of gravitational instabilities on long-term binary orbital evolution has been examined in, e.g., \cite{Franchini2021, Bortolas2021}. 
Although \cite{Franchini2021} only simulated equal mass binaries and \cite{Bortolas2021} only studied $q \gtrsim 0.1$, self-gravity is not likely to influence the EM signatures of the late inspiral regardless of binary mass ratio.
\begin{figure}
    \centering
    \includegraphics{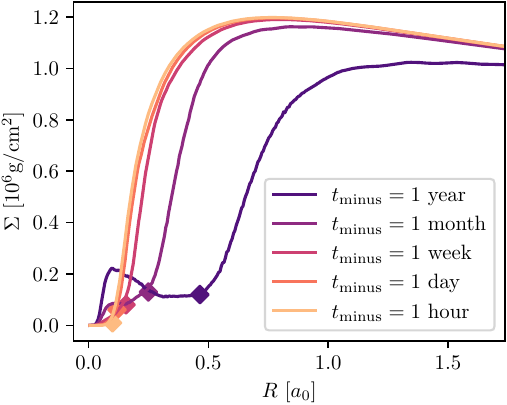}
    \caption{The radial profile of the surface density $\Sigma(r)$ in a binary with mass ratio $q = 0.01$ and orbital Mach number $\mathcal{M} = 10$, shown at representative times before the merger. The radial profiles are obtained from the azimuthal average of the two-dimensional disk surface density. Diamond markers show the location of the secondary at the representative times.}
    \label{fig:SigmaProfile}
\end{figure}
\begin{figure}
    \centering
    \includegraphics{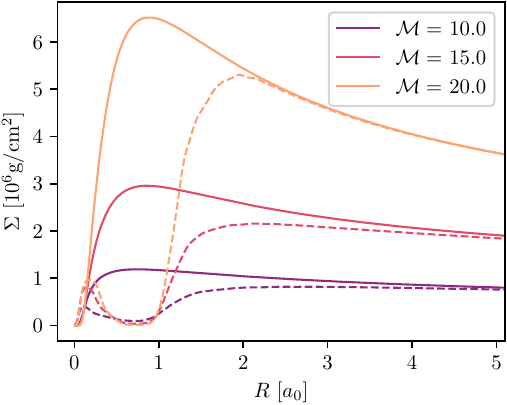}
    \caption{The radial profile of the surface density $\Sigma(r)$ in a binary with mass ratio $q = 0.01$ and orbital Mach numbers in the range $\mathcal{M} = 10, 15, 20$, shown at $4000$ orbits before merger (dashed) and at merger (solid). The radial profiles are obtained from the azimuthal average of the two-dimensional disk surface density.}
    \label{fig:Sigma-diffMach}
\end{figure}

\section{Numerical Methods}
\label{sec:numerical}
Our hydrodynamic simulations are performed using the publicly available \texttt{Sailfish} code. \texttt{Sailfish} is a GPU-accelerated, grid-based hydrodynamics code with physics and post-processing features specifically targeting problems related to binary-disk interactions. \texttt{Sailfish} uses a fixed-mesh, 2nd-order Godunov solver and has been used in many published studies of binary accretion \citep[e.g.][]{WesternacherSchneider2022, WesternacherSchneider2023, Krauth2023a, Krauth2023b, Duffell2024, Tiede2024a, DeLaurentiis2024}.

\subsection{Equations of Motion}
\label{subsec:Equations}
Our simulations are based on solutions to the vertically averaged, time-dependent mass continuity and Navier-Stokes equations,
\begin{align}
    \label{eqn:NS1}
    &\frac{\partial \Sigma}{\partial t} + \mathbf{\nabla}\cdot(\Sigma\mathbf{v}) = \dot{\Sigma}_{\rm sink} \, , \\
    &\frac{\partial \Sigma \mathbf{v}}{\partial t} + \mathbf{\nabla } \cdot (\Sigma \mathbf{v}\mathbf{v} + P\,\mathbf{I} - \mathbf{T}_{\text{vis}}) = \dot{\Sigma}_{\rm sink} \mathbf{v} + \mathbf{F}_g \, .
    \label{eqn:NS2}
\end{align}
Here, $\Sigma$ is the vertically-integrated mass density, $\mathbf{v}$ is the gas velocity, $P \equiv \Sigma c_s^2$ is the vertically-integrated gas pressure, and $\mathbf{I}$ is the identity tensor (pressure is isotropic). We adopt a local isothermal equation of state, $c_s^2 = -\mathbb{V} / \mathcal{M}^2$, where $\mathcal{M}$ is the orbital Mach number, and the gravitational potential is given by
\begin{align}
    \mathbb{V} = -\frac{GM_1}{(r_1^2 + r_s^2)^{1/2}} - \frac{GM_2}{(r_2^2 + r_s^2)^{1/2}} \ .
    \label{eqn:potential}
\end{align}
Here $r_1$ and $r_2$ are the distances to each black hole, and $r_s$ is the gravitational softening length to ensure the potential is finite at the component positions. Use of the local isothermal equation of state means that we do not solve an energy transport equation nor adopt a phenomenological cooling prescription. Certain thermodynamic processes, such as shock-heating, are not accounted for as a result of the isothermal equation of state.

The ``sink'' term $\dot{\Sigma}_{\rm sink}$, appearing in Eqns. \ref{eqn:NS1} and \ref{eqn:NS2}, is responsible for the direct exchange of mass and momentum between the gas and the binary components,
\begin{align}
    \dot{\Sigma}_{\rm sink} = -\frac{\Sigma} {\tau_{\rm sink}} \bigg( e^{-r_1^2\, /\, 2r_{\rm sink}^2} + e^{-r_2^2\, /\, 2r_{\rm sink}^2} \bigg) \ .
    \label{eqn:sink}
\end{align}
We have checked that our choice of sink rate, $\tau_{\rm sink}^{-1}$, does not significantly affect our measurements of the preferential accretion rate, provided the sink rate is sufficiently large. This seems to be different from \cite{Dittmann2024}, who reported modest sensitivity to the sink rate. We suspect the difference could be in the use of $\alpha$ versus constant-$\nu$ viscosity. In this study, we have adopted a sink radius of $r_{\rm sink} = 0.05 a_0$ and a sink rate of $\tau_{\rm sink}^{-1} = 10 \Omega_0$.

Our chosen sink radius is always smaller than the Hill radius of the secondary black hole, $r_{\rm hill} = a_0 (q / 3)^{1/3} \ge 0.07a_0$, at the start of the simulation, even for the lowest mass ratio considered ($q = 10^{-3}$). However, as the binary separation decreases, the sink size of $0.05a_0$ will become too large to resolve the Hill radius for lower mass ratios. We acknowledge this as a caveat, as our results may be sensitive to sink size, particularly for the smallest mass ratios simulated. To address this issue, we distinguish our accretion rate results based on the sink radius of the secondary black hole. When the sink radius satisfies $r_{\rm sink} \ge 0.5 \ r_{\rm hill}$, we represent the results with dashed, less opaque lines to indicate that gas flows in the circum-secondary disk are not well resolved. For a discussion of the sensitivity of the accretion rates to the sink size and sink rate in our simulations, see Appendix \ref{sec:appendix}.

The vertically integrated gravitational force density, associated with the softened gravitational potential, is $\mathbf{F}_g = -\Sigma\,\nabla\mathbb{V}$. We assume the binary mass $M = M_1 + M_2$ is much larger than the disk mass so that self-gravity can be safely ignored (see Sec. \ref{subsec:assumptions}). The components of the viscous stress tensor $\mathbf{T}_{\text{vis}}$ in Eqn. \ref{eqn:NS2} are given by $T^{ij}_{\text{vis}} = \nu \Sigma \left(\partial^j v^i + \partial^i v^j - \partial_k v^k \delta^{ij} \right)$. \texttt{Sailfish} supports both ``constant-$\nu$'' and ``constant-$\alpha$'' viscosity prescriptions. For this study, we have adopted the constant-$\alpha$ viscosity prescription with kinematic viscosity coefficient
\begin{equation}
    \nu =
    \frac{\alpha c_s^2}{\tilde{\Omega}} \, ,
    \label{eq:alphavisc}
\end{equation}
where the modified Keplerian orbital frequency $\tilde{\Omega}$ is
\begin{equation}
    \tilde{\Omega} = \sqrt{
    \frac{GM_1}{r_1^3} + \frac{GM_2}{r_2^3} }\, .
    \label{eq:Omega}
\end{equation}
At large radii $r \gg a$, Eqn. \ref{eq:alphavisc} reduces to $\nu \simeq \bar{\nu} \sqrt{GMr}$ where $\bar{\nu} = \alpha / \mathcal{M}^2$.

The black holes are moved according to the Kepler two-body problem, modified to account for the GW-driven inspiral up to the leading post-Newtonian order \citep{Peters1964}. The semi-major axis as a function of time is the solution of Eqn. \ref{eq:adot},
\begin{equation}
    a(t) = a_0 \left(1 - 4 t / \tau_0 \right)^{1/4} \, ,
    \label{eq:a(t)}
\end{equation}
where $\tau_0$ is the contraction timescale at the start of the simulation. Positioning the binary components on the grid requires an expression for the orbital phase $\phi(t)$ as a function of the simulation time $t$. The orbital phase is obtained by integrating in time the instantaneous orbital frequency,
\begin{equation}
   \phi(t) = \int_{a_0}^{a(t)} \sqrt{\frac{GM}{a^3}} \frac{da}{\dot a_{\rm gw}} = 
   \frac{2}{5} \tau_0 \Omega_0 \left(1 - \left(\frac{a(t)}{a_0}\right)^{5/2}\right)
   \, .
   \label{eq:phi}
\end{equation} 
The gravitational force of the gas is neglected in the binary equation of motion, which is justified because we are only considering binaries deep in the GW-driven regime.

\begin{figure}
    \centering
    \includegraphics{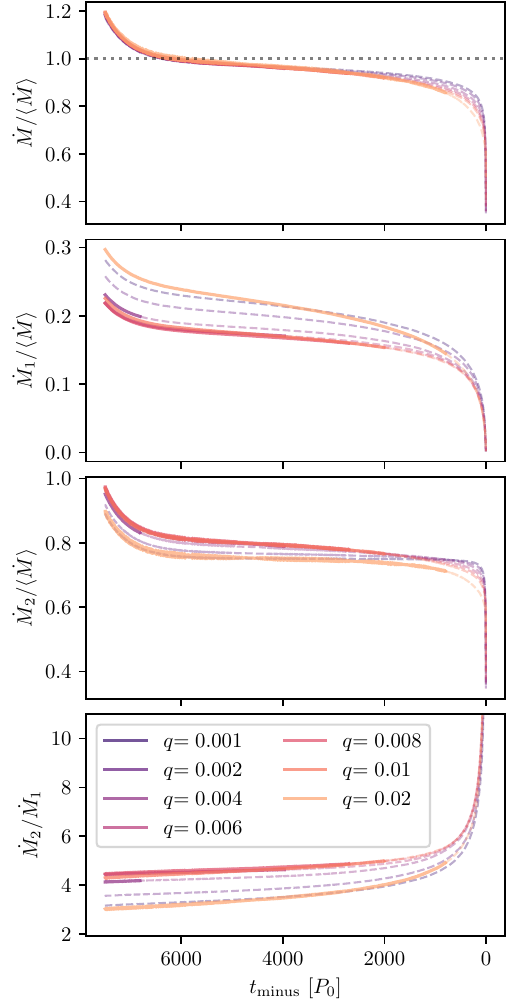}    \caption{Time series of the mass accretion rates $\dot{M}$ (top), $\dot{M}_1$ (second), $\dot{M}_2$ (third), and the preferential accretion parameter $\dot{M}_2 / \dot{M}_1$ (bottom), for mass ratios $q$ in the range of $0.001 - 0.02$. Each line starts as a solid line and transitions to a dashed line when the sink radius of the secondary BH becomes comparable to the size of the Hill radius, indicating that gas flows in the secondary disk are not well resolved for the dashed curves. The gas orbital Mach number is $\mathcal{M} = 10$. The dotted line in the top panel indicates where $\dot{M} = \langle \dot{M} \rangle$.}
    \label{fig:Mdot}
\end{figure}
\begin{figure}
    \centering
    \includegraphics{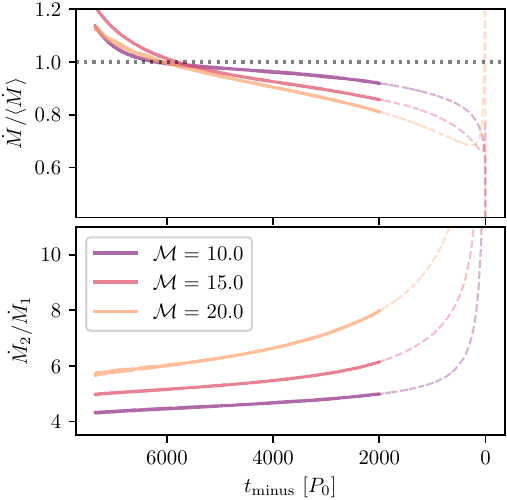}
    \caption{Time series of the mass accretion rates $\dot{M}$ (top) and the preferential accretion parameters $\dot{M}_2 / \dot{M}_1$ (bottom) for a range of Mach numbers, $\mathcal{M} = 10, 15, 20$. The binary mass ratio is $q = 0.01$. Each curve in the top panel is normalized by the empirical average binary accretion rate throughout the inspiral, $\langle \dot{M} \rangle$. Each line starts as a solid line and transitions to a dashed line when the sink radius of the secondary BH becomes comparable to the size of the Hill radius, indicating that gas flows in the secondary disk are not well resolved for the dashed curves. The dotted line in the top panel indicates where $\dot{M} = \langle \dot{M} \rangle$.}
    \label{fig:Mdot-diff_mach}
\end{figure}

\subsection{Initial Data, Outer Boundary Condition, and Diagnostics}
\label{subsec:initial}
We start our simulations of accretion onto the binary with a circular gas disk: the azimuthal velocity profile is $\varv_\phi = \varv_{\rm kep} = \sqrt{G M / r}$, the radial velocity is inward at the viscous drift speed, $\varv_r = -3 \nu(r) / 2r$, and the surface density is written
\begin{equation}
    \Sigma_0(r) = \frac{\dot M_0}{3 \pi \nu(r)} \propto r^{-1/2} \, ,
    \label{eqn:steady-state-sigma-zero-torque}
\end{equation}
generally as a steady-state viscous disk in the middle and on the right-hand side approximated for an $\alpha$-disk with viscosity as in Eqn. \ref{eq:alphavisc} and with uniform orbital Mach number. Note that Eqn. \ref{eqn:steady-state-sigma-zero-torque} is the surface density of a steady disk with a zero net angular momentum current; this fact is revisited in Sec. \ref{subsec:steadystate}. After the start of the simulation, it takes several viscous timescales at $r \sim a_0$ for the flow to relax to a statistically steady state. For reference, when $\alpha = 0.05$ and $\mathcal{M} = 10$, the viscous timescale at $r=a_0$ is $t_{\rm visc} \simeq 200 \ {\rm orbits}$. For both a $q=0.01$ and $q=0.1$ binary, Fig. \ref{fig:sailfish} shows a snapshot of the surface density once the system has viscously relaxed to a quasi-steady state, or in other words, the surface density snapshot is taken after a thousand orbits or five viscous timescales at $r=a_0$. 

Because our code uses Cartesian coordinates with a square-shaped domain, we impose a ``soft'' outer boundary condition within a circular buffer zone starting at radius $r_{\rm buf} \simeq 7.5 a_0$ (we have confirmed that our results are not changed by an increase of $r_{\rm buf}$, see Sec. \ref{subsec:steadystate}). Inside the buffer region, the surface density and gas velocity are driven smoothly toward a zero-torque $\alpha$-disk (which is also the initial condition). The domain size of our simulations is $20 a_0$, and the number of grid zones is $2400 \times 2400$ unless mentioned otherwise; the simulations considered in this paper have a grid spacing $\Delta x \simeq 0.008 a_0$. 

The accretion rate for each black hole is measured by separately integrating the two terms in Eqn. \ref{eqn:sink} corresponding to the primary and secondary black hole accretion rates. Smoothing is applied to the time series of the accretion rates unless specifically mentioned. The time series of the gravitational torque is computed by integrating the gravitational torque density over the whole surface of the disk,
\begin{equation}
    \dot J_{\rm grav} = \int r \times \mathbf{F}_g \ dA \, .
    \label{eq:Jdot_grav}
\end{equation}
The gravitational torque is decomposed into the torque on the primary or the secondary and also into contributions from the inner and outer disks. We ignore the accreted spin angular momentum as we have calculated that the accretion torques are negligible.

\subsection{A Note About Accretion Rates in the Steady State}
\label{subsec:steadystate}
Ideally, our accretion rates $\dot M_1$ and $\dot M_2$ would be reported relative to the theoretical inflow rate $\dot M_0$ in Eqn. \ref{eqn:steady-state-sigma-zero-torque}. However, we have observed that in our simulations, the mean binary accretion rate $\langle \dot M \rangle$ generally exceeds $\dot M_0$ by a significant margin. We understand this ``overshoot'' to be an artifact of our treatment of the domain outer boundary (Sec. \ref{subsec:initial}).

Near the outer boundary (or the buffer region), the surface density is driven smoothly toward $\Sigma_0(r)$ in Eqn. \ref{eqn:steady-state-sigma-zero-torque}, which corresponds to a zero-torque disk. We make this choice mainly for convenience because the torque on the binary $\dot J$ cannot be controlled. Rather, it needs to be measured from the simulation. When $\dot J$ is positive, a better approximation of the surface density far from the binary is \citep[see e.g.][]{Pringle1981}
\begin{equation}
    \Sigma(r) = \frac{1}{3 \pi \nu(r)} \left(\langle \dot M \rangle - \frac{\dot J}{\sqrt{G M r}} \right) \, .
    \label{eqn:steady-state-sigma}
\end{equation}
Thus the target surface density $\Sigma_0(r)$ would be generally smaller than $\Sigma(r)$ if we had $\langle \dot M \rangle = \dot M_0$. However the buffer zone forces the equality $\Sigma(r_{\rm buf}) = \Sigma_0(r_{\rm buf})$, so in general we have $\langle \dot M \rangle \gtrsim \dot M_0$.

The amount of this ``overshoot'' can be estimated, noting that when the binary mass ratio is small, the gravitational torque is negligible, and the net angular momentum current is then $\dot J \simeq \dot M_2 \Omega a^2$. We check if the gravitational torque is negligible in Sec. \ref{subsec:torques} and find that for a mass ratio of $q = 0.01$, the gravitational torque is non-negligible. However, as an approximation, we find that the ratio of the measured inflow rate $\langle \dot M \rangle$ to the nominal inflow rate $\dot M_0$ is then
\begin{equation}
   \frac{\langle \dot M \rangle}{\dot M_0} \simeq \frac{1}{1 - \sqrt{a / r_{\rm buf}}} \, .
   \label{eqn:mdot-overshoot}
\end{equation}
In Eqn. \ref{eqn:mdot-overshoot} we have also assumed the secondary dominates a majority of the accretion, i.e. $\dot M_2 \simeq \langle \dot M \rangle$. In simulations where $r_{\rm buf} = 7.5 a_0$, Eqn. \ref{eqn:mdot-overshoot} predicts $\langle \dot M \rangle / \dot M_0 \simeq 1.57$, which is very close to what we observe.

\begin{table*}
    \centering
    \begin{tabular}{r|l}
         Binary Parameter &  Initial Value \\
         \hline 
         \hline
         Initial separation: & $a_0 = 3.6 \ \rm{AU} \ \bigg(\frac{M}{10^7 \rm{M}_\odot}\bigg) \bigg(\frac{\eta}{0.01}\bigg)^{2/5} \bigg(\frac{\tau/P}{10^4 \ \rm{orbits}}\bigg)^{2/5}$ \\
         \hline
         Initial time until merger: & $t_{\rm{minus}} =  5.4 \ \rm{years} \ \bigg(\frac{M}{10^7 \rm{M}_\odot}\bigg)^{-1/2} 
         \bigg(\frac{a_0}{3.6 \ \rm{AU}}\bigg)^{3/2}
         \bigg(\frac{\tau/P}{10^4 \ \rm{orbits}}\bigg)$\\
         \hline
         Initial primary BH gravitational radius: & $r_{g_1} = 2.7 \times 10^{-2} \ a_0 \ \bigg(\frac{M}{10^7 \rm{M}_\odot}\bigg) \bigg(\frac{1+q}{1.01}\bigg)^{-1} $ \\
         \hline
         Initial secondary BH gravitational radius: & $r_{g_2} = 2.7 \times 10^{-4} \ a_0 \ \bigg(\frac{M}{10^7 \rm{M}_\odot}\bigg) \bigg(\frac{\eta}{0.01}\bigg) $
    \end{tabular}
   \caption{Initial values for the binary parameters used in the simulation results presented in this manuscript. These values are based on the fiducial binary system with $q = 0.01$, $M = 10^7 \ \rm{M}_{\odot}$, and $10^4$ orbits until merger, but they can be scaled accordingly for any binary black hole inspiral.}
    \label{tab:initial-values}
\end{table*}

We have checked that the overshoot does not contaminate any of our derived science results, such as the shape of the accretion rate time series. Overshoots of the kind we observed here could be mitigated by guessing the value of $\dot J / \dot M$ and then driving the solution in the buffer region toward $\Sigma(r)$ in Eqn. \ref{eqn:steady-state-sigma}. Note that overshoots of the binary accretion rate have also been reported in \cite{Farris2014} and \cite{Shi2015} and were also discussed in \cite{Rafikov2016}. We suggest here that the finite domain size could have been responsible for those observations.

Note, on the other hand, the accretion overshoot due to the finite outer boundary introduced in this section is not the same as the accretion overshoot described in \cite{Miranda2017}, \cite{Dittmann2022}, and \cite{Duffell2024}. The accretion overshoot described in those works will, over many orbits, slowly approach the theoretical inflow rate $\dot M_0$ as the disk settles to larger radii. The lasting overshoot we observe here is due to the finite radius of the domain outer boundary, and the gradual decline of the accretion rate toward $\dot M_0$ is halted when the disk would need to be viscously relaxed at radii larger than $r_{\rm buf}$.

\section{Simulation Results}
\label{sec:results}
We simulated the GW-driven inspirals of black hole binaries with mass ratios between $q=10^{-3}$ and $q = 10^{-1}$, surrounded by disks with orbital Mach numbers $\mathcal{M} = 10, 15, 20$. The binary mass is $M = 10^7 M_\odot$; however, the accretion rate time series could be rescaled by an appropriate remapping of the time coordinate to describe a binary of any mass. Results related to observer time scales, EM light curves, and emission spectra depend on $M$, but they also scale with it in obvious ways. Simulations generally begin on the order of $\tau / P \sim 10^4$ binary orbits prior to the coalescence. In Sec. \ref{subsec:numerical}, we show that this provides sufficient time for the disk to relax and ``forget'' the simulation's initial condition well before the viscous decoupling stage.
Times are measured in the binary rest-frame, not in the observer time, i.e., we do not adjust the time scales for cosmological redshift.

We have summarized the initial binary parameters used in our results in Table \ref{tab:initial-values} for a binary with a total mass of $M = 10^7 \ \rm{M}_{\odot}$, a mass ratio of $q = 0.01$, and $10^4$ orbits until merger. As shown in the table, these values can be scaled for different total binary masses, mass ratios, and numbers of orbits before merger using Eqns. \ref{eq:adot} and \ref{eq:notm-rg-v}. Consequently, not all results in this work use the exact initial values given in Table \ref{tab:initial-values}, as our study explores a broad range of mass ratios and varying initial orbits before merger.

\subsection{Evolution of Disk Surface Density}
\label{subsec:sigma}
Fig. \ref{fig:SigmaProfile} shows the surface density radial profiles of the fiducial disk, initially at $t_{\rm minus}=1$ year and then at four other representative times leading up to merger. At radii $r \gtrsim a_0$, the surface density profile is approximately $\Sigma \propto r^{-1/2}$, as for a steady state $\alpha$-disk in Eqn. \ref{eqn:steady-state-sigma-zero-torque}. The surface density evolution is similar in certain ways to the one-dimensional disk model from \cite{Chang2010}. An annular gap is formed around the secondary orbit, and the outer disk follows the secondary inwards up until the viscous decoupling time. However, the gap in our simulation is not fully evacuated; gas flows across the gap and accretes to the secondary. Also, the surface density of the inner disk in our simulation is seen to be modestly decreasing over time, whereas in \cite{Chang2010}, the presence of the vacuum gap around the secondary orbit requires the surface density of the inner disk to increase. We show in the next sub-section that the decrease of the inner disk mass is driven mainly by accretion to the secondary. In other words,  negligible mass passes the secondary's orbit, and instead, most gas is captured by the secondary, gradually starving the primary.

The surface density radial profiles for varying Mach number disks are shown in Fig. \ref{fig:Sigma-diffMach} at four thousand orbits before merger (dashed) and at merger (solid). 
As the Mach number increases, the viscosity decreases, and while the inflow rate $\dot{M}_0 = 3\pi\nu\Sigma$ is held constant across all runs, the surface density magnitude must then increase. This explains the shift in surface density for higher Mach number disks.

\subsection{Long-Term Viscous Decoupling}
\label{subsec:mdots}
Fig. \ref{fig:Mdot} shows the binary accretion rate $\dot M$ (top), the accretion rates to the binary components $\dot M_1$ and $\dot M_2$ (middle), and the preferential accretion rate $\dot{M}_2 / \dot{M}_1$ (bottom). The curves show the time series data for binaries with various mass ratios. Our main results here can be summarized, (1) the secondary generally starves more slowly than the primary, and (2) the binary accretion rate $\dot M$ shows a long-term downward trend, being reduced from $\langle \dot M \rangle$ by $10\%$ as early as $\sim\! 500 \tau_{\rm dec}$ (decade-like time span) before the merger.

The second result is surprising, as one could reasonably guess that $\dot M$ would be significantly reduced from $\dot M_0$ only within a few $\tau_{\rm dec}$ of the merger, perhaps following a functional form like $\dot M = \dot M_0 e^{-\tau_{\rm dec} / t_{\rm minus}}$. Instead, our results suggest the quantity $\delta \equiv 1 - \dot M / \dot M_0$ might grow as a power law in $t_{\rm minus}$, at least while $\delta$ is small. We summarize next a model of the viscous decoupling process, which accounts for the time evolution of the binary-disk torque and predicts such scaling of $\delta$. A full derivation is given in \cite{Zrake2025}.


%
\begin{figure}
    \centering
    \includegraphics{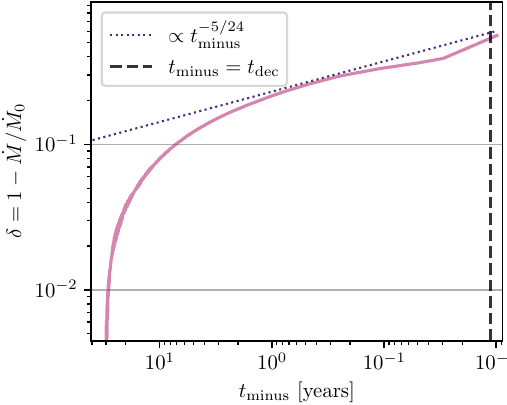}
    \caption{Time series of the $\delta = 1 - \dot M / \dot M_0$ parameter for the fiducial binary with mass ratio $q=0.01$ and $\mathcal{M}=10$. The dotted line represents the predicted power-law dependence, and the dashed vertical line is the decoupling timescale.}
    \label{fig:delta}
\end{figure}

Our model characterizes the disk as having two distinct zones: a close zone that is tightly coupled to the binary and a distant zone that is relatively frozen over the remaining time before the merger. The two zones meet at a viscous radius, $r_\nu$, satisfying $t_{\rm visc}(r_\nu) = -a / \dot a_{\rm gw}$ \footnote{This idea is inspired by Sec. 3.2 of \cite{Rafikov2016}, which defines a viscous radius, but in a different setting where the binary orbit is held fixed while the disk viscously adjusts to a new angular momentum current. The idea is also used in Sec. 5.6 of \cite{Duffell2024} to explain the damping of a startup transient in simulations of binary accretion initiated with a zero-torque disk.}. In the close zone $r < r_\nu$, the disk evolves through a sequence of steady states (Eqn. \ref{eqn:steady-state-sigma}), each conducting angular momentum inward at rate $\dot J = \dot M \Omega a^2$. At each step $n$ of the sequence, a boundary condition is imposed, $\Sigma_n(r_n) = \Sigma_{n-1}(r_n)$, where $r_n$ is the viscous radius at step $n$. Such recursion leads to a first-order ordinary differential equation for $\delta$, and the solutions go asymptotically as $\delta \propto t_{\rm minus}^{-5/24}$ \citep{Zrake2025}. Such weak dependence of $\delta$ on $t_{\rm minus}$ could account for the long-term viscous decoupling seen in our simulations. Fig. \ref{fig:delta} shows a plot of $\delta(t)$ for the fiducial binary compared to the asymptotic prediction $\delta \propto t_{\rm minus}^{-5/24}$. With many decoupling timescales before the merger, the simulations agree with the predicted power-law dependence.

In Sec. \ref{subsec:dec}, we estimated that the viscous decoupling timescale would be very sensitive to the Mach number of the disk, $\tau_{\rm dec} \propto \mathcal{M}^{16/5}$. However, given our argument above, where $\delta$ scales as a power-law in $t_{\rm minus}$, it seems reasonable to expect that the late-inspiral accretion rate curves might have only modest sensitivity to the Mach number. To test this, we ran a suite of simulations with a range of Mach numbers, $\mathcal{M} = 10, 15, 20$. The results appear in Fig \ref{fig:Mdot-diff_mach}, and do confirm this expectation. The run with Mach $20$ has a nominal viscous decoupling time roughly $10$ times larger than the run with Mach $10$, and yet the curves differ mainly by a vertical offset. The long-term decoupling effect seen here also seems to be present in simulations of equal-mass inspirals in, e.g., Fig. 6 in \cite{Farris2014}, or Fig. 3 of \cite{Dittmann2023}.

\begin{figure}
    \centering
    \includegraphics{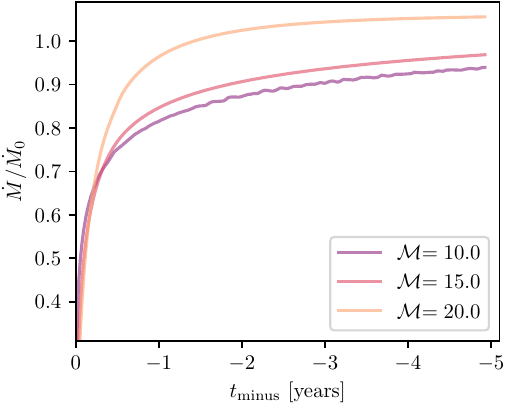}
    \caption{Time series of the post-merger mass accretion rate for the fiducial model, but with a range of Mach numbers $\mathcal{M} = 10, 15, 20$.}
    \label{fig:Mdot-refill}
\end{figure}

\subsection{Remnant Fueling}
\label{subsec:remnant}
Equally important to the pre-merger accretion evolution is the post-merger remnant refueling, where the disk viscously refills the cavity and accretes onto the merger remnant. In Sec. \ref{subsec:viscous-closing}, we derived the refilling timescale $t_{\rm refill}$ (Eqn. \ref{eq:t_refill}) and found that for the fiducial binary, the disk should refill the cavity, and begin accreting to the black hole merger remnant roughly $4$ days after the merger. Note, however, that our estimate predicts $t_{\rm refill} \propto \mathcal{M}^{16/5}$, analogous to the viscous decoupling time scale $\tau_{\rm dec}$. It would imply that the onset of refueling could be delayed from days (Mach $10$) to decades if the disk had a Mach number of $100$.

In Fig. \ref{fig:Mdot-refill}, we plot the total binary accretion rate post-merger out to $5$ years after coalescence for a binary with mass ratio $q=0.01$ and for simulations with different Mach numbers. The result indicates that the onset of refueling happens over a time span of months to years, with relatively little sensitivity to our derived estimate for $t_{\rm refill}$.
We thus suggest that the refilling timescale post-merger is not well represented by a time constant such as that given in Eqn. \ref{eq:t_refill}, as the recovery of the accretion rate to $\dot M_0$ does not show any exponential characteristics. The exact functional form of the refueling curve can be predicted from the model, which we summarized in Sec. \ref{subsec:mdots}, and comparisons are presented in \cite{Zrake2025}.

\subsection{Variability of Mass Accretion Rates}
\label{subsec:variability}

Low-mass ratio binaries seem to accrete with lower stochastic variability than binaries of near-equal mass \citep[e.g.][]{Farris2014, Shi2015, DOrazio2016, Duffell2020, Dittmann2024}. This suggests that accreting IMRI\ap{s} could produce ``cleaner'' light curves, possibly making it easier to identify them based on EM periodicity, or color evolution, in the years before a merger\footnote{Note a caveat of this prediction is that AGN tends to, in general, be highly variable. Therefore any variability due to a binary could be difficult to detect regardless of binary mass ratio. Additionally, \cite{Noble2012} discovered using 3D GRMHD simulations that the luminosity has a suppressed variability amplitude relative to accretion rates, suggesting that our predictions could be optimistic.}. In this section, we show that the reduced stochasticity seen by \cite{Duffell2020} and \cite{Dittmann2024} applies throughout a GW-driven inspiral.

\begin{figure}
    \centering
    \includegraphics{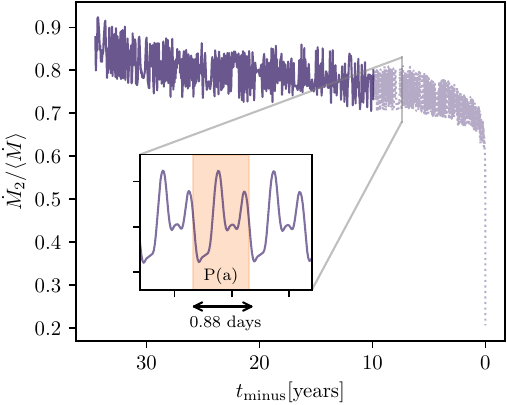}
    \caption{Time series of the secondary black hole mass accretion rate for a binary with a mass ratio of $q = 0.01$ and $\mathcal{M} = 10$. The accretion rate curve starts as a solid line and transitions to a dashed line when the sink radius of the secondary BH becomes comparable to the size of the Hill radius, indicating that gas flows in the secondary disk are not well resolved for the dashed curves. The highlighted region in the inset plot shows the (unsmoothed) accretion rate during one rest-frame binary orbit, which is approximately $\unit[0.9]{days}$ at $t_{\rm minus} = \unit[8]{years}$ when $q = 0.01$.}
    \label{fig:Mdot2insetq0.01}
\end{figure}

\begin{figure}
    \centering
    \includegraphics{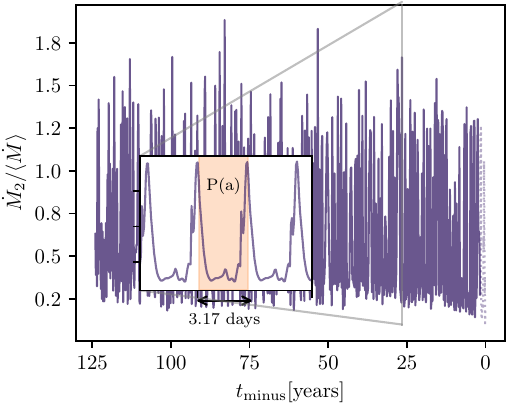}
    \caption{Time series of the secondary black hole mass accretion rate for a binary with a mass ratio of $q = 0.1$ and $\mathcal{M} = 10$. The accretion rate curve starts as a solid line and transitions to a dashed line when the sink radius of the secondary BH becomes comparable to the size of the Hill radius, indicating that gas flows in the secondary disk are not well resolved for the dashed curves. The highlighted region in the inset plot shows the (unsmoothed) accretion rate during one rest-frame binary orbit, which is approximately $\unit[3.2]{days}$ at $t_{\rm minus} = \unit[25]{years}$ when $q = 0.1$.}
    \label{fig:Mdot2insetq0.1}
\end{figure}

The time series of the secondary accretion rate $\dot M_2$ is shown in Fig. \ref{fig:Mdot2insetq0.01} for a $q=0.01$ binary, and in Fig. \ref{fig:Mdot2insetq0.1} for a $q=0.1$ binary. The amplitude of the stochastic variability in $\dot M_2$ is less than roughly $10\%$ for the case of $q=0.01$ and greater than roughly $50\%$ for the case of $q=0.1$, which confirms there is a significant increase in the variability amplitude somewhere between $q=0.01$ and $q=0.1$, and that this change persists also in the late inspiral phase, at least before the viscous decoupling.

The periodic variability of unequal-mass systems can be seen in the insets of Figs. \ref{fig:Mdot2insetq0.01} and \ref{fig:Mdot2insetq0.1}. The secondary accretion rate modulates at the binary orbital period, and we believe this could indicate a slight eccentricity of the outer disk \citep[e.g.][]{Kley2006}. For lower mass ratios, a second peak is more predominant in the secondary's accretion rate. This half-orbital periodicity has been seen in other works \citep[e.g.][]{Farris2014, Dittmann2024, Cocchiararo2024}, and is likely the signature of an eccentric outer disk \footnote{\cite{Cocchiararo2024} also reported a second peak in the total accretion rate; the minor peak we see in the secondary accretion rate seems compatible with their results.}. We have measured the eccentricity of the outer disk in our runs by summing the local gas eccentricity vectors and found the eccentricity to be approximately $2 \%$ for the $q=0.01$ binary and around $14 \%$ for the $q=0.1$ binary.


\subsection{Evolution of Binary Torques}
\label{subsec:torques}
\begin{figure}
    \centering
    \includegraphics{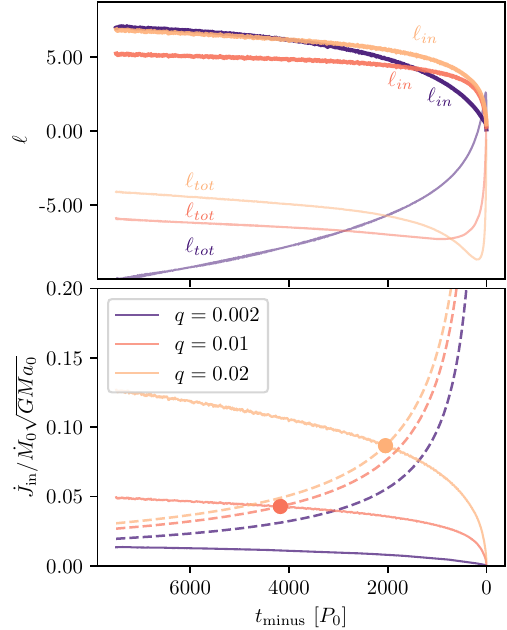}
    \caption{Evolution of the gravitational torque, normalized in different ways, for mass ratios $q = 0.002, 0.01, 0.02$ and Mach number $\mathcal{M} = 10$. In the bottom panel, the solid lines are the measured torque $\dot J_{\rm in}$, applied to the binary by the inner disk. The dashed lines show the minimum torque $\dot{J}_{\rm plow}$ (Eqn. \ref{eq:Jdotplow}) needed to sustain tidal squeezing, and the filled circles are where $\dot{J}_{\rm in} = \dot{J}_{\rm plow}$.}
    \label{fig:jdotin}
\end{figure}
\begin{figure}
    \centering
    \includegraphics{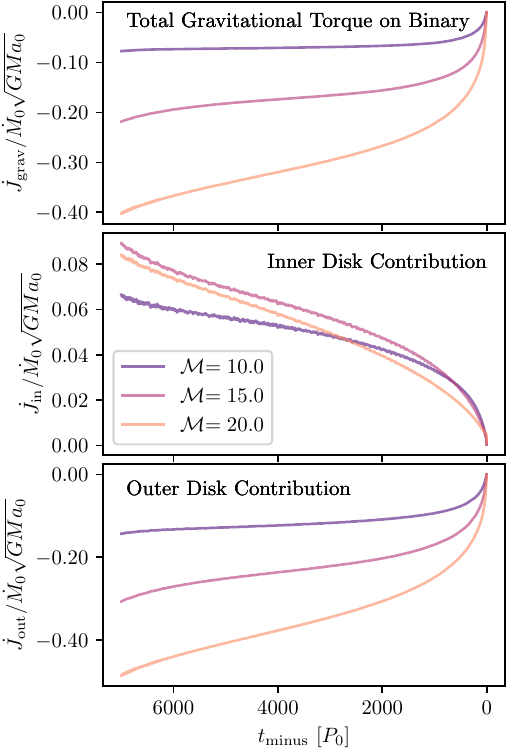}
    \caption{Evolution of the total gravitational torque $\dot J_{\rm grav}$ (top), inner disk gravitational torque $\dot J_{\rm in}$ (middle) and the outer disk gravitational torque $\dot J_{\rm out}$ (bottom) on the binary. The mass ratio is $q = 0.01$ and the Mach numbers are $\mathcal{M} = 10, 15, 20$.}
    \label{fig:jdot-diffMach}
\end{figure}
\begin{figure}
    \centering
    \includegraphics{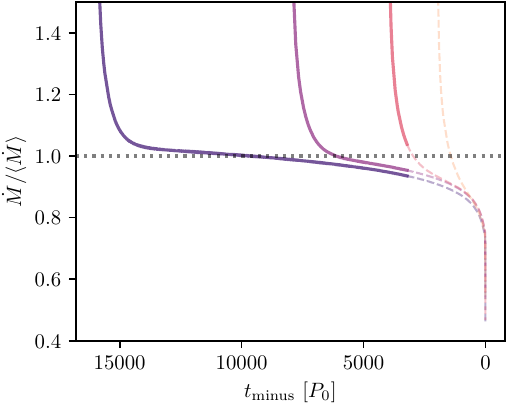}
    \caption{Time series of the binary mass accretion rate for a mass ratio of $q = 0.01$ and an orbital Mach number of $\mathcal{M} = 10$. The curves are from distinct runs initiated progressively earlier in the inspiral. The dotted line indicates where $\dot{M} = \langle \dot{M} \rangle$. Each curve starts as a solid line and transitions to a dashed line when the sink radius of the secondary BH becomes comparable to the size of the Hill radius, indicating that gas flows in the secondary disk are not well resolved for the dashed curves.}
    \label{fig:Mdot-diff_tstart}
\end{figure}

Fig. \ref{fig:jdotin} shows the time series of the gravitational torque on the binary for three mass ratios ($q = 0.002, 0.01, 0.02$). The top panel shows the non-dimensional torques $\ell_{\rm tot} = (\dot{J}_{\rm tot}/J)/(\dot{M}/M)$ and $\ell_{\rm in} = (\dot{J}_{\rm in}/J)/(\dot{M}/M)$, and the bottom panel shows the inner disk gravitational torque $\dot J_{\rm in}$, normalized by $\dot M_0 \sqrt{G M a_0}$. Fig. \ref{fig:jdot-diffMach} shows the gravitational torque on the binary for simulations of varying Mach numbers. The top panel shows the total gravitational torque exerted on the binary, and the middle and bottom panels show, respectively, the inner and outer disk contributions. 

In the top panel of Fig. \ref{fig:jdot-diffMach}, we find that the net gravitational torque on the binary is negative for all of our runs and that it becomes increasingly negative with increasing Mach number. This trend is consistent with other studies, including \cite{Tiede2020} and \cite{Penzlin2022}, which also find that the torque on the binary becomes increasingly negative with increasing Mach number, albeit these studies focused on equal mass or high mass ratio binaries. \cite{Dittmann2024}, on the other hand, simulated unequal mass binaries down to $q=0.01$ and measured the gravitational torque of a $q=0.01$ binary with $\mathcal{M}=10$ to be marginally positive with $\dot{J}_{\rm grav} \simeq 0.1$ and no clear trend for varying Mach number. Our results in Fig. \ref{fig:jdot-diffMach}, for the same binary system, show a marginally negative total gravitational torque of $\dot{J}_{\rm grav} \simeq 0.1$ with a clear trend of monotonically decreasing torque with increasing Mach number. We believe the (small) difference might be attributable either to the difference in the viscosity prescription (constant-$\alpha$ vs. constant-$\nu$) or differences in the sink rate, as \cite{Dittmann2024} found some sensitivity of their torques to the value selected for the sink rate.

The outer disk is responsible for the increasingly negative torque at a higher Mach number; the inner disk actually exerts a slightly positive torque on the binary, which is rather insensitive to the Mach number. The increasing magnitude of torque from the outer disk can be attributed to the accumulation of mass outside the secondary orbit (Fig. \ref{fig:Sigma-diffMach}). 

These results could be used to predict how the gravitational waveform of IMRI\ap{s} surrounded by gas should differ from those evolving in vacuum \citep[e.g.][]{Derdzinski2019,Derdzinski2021,Tiede2024b}. Such predictions will be reported in a future study focusing specifically on GW signatures.

\begin{figure*}
    \centering
    \includegraphics{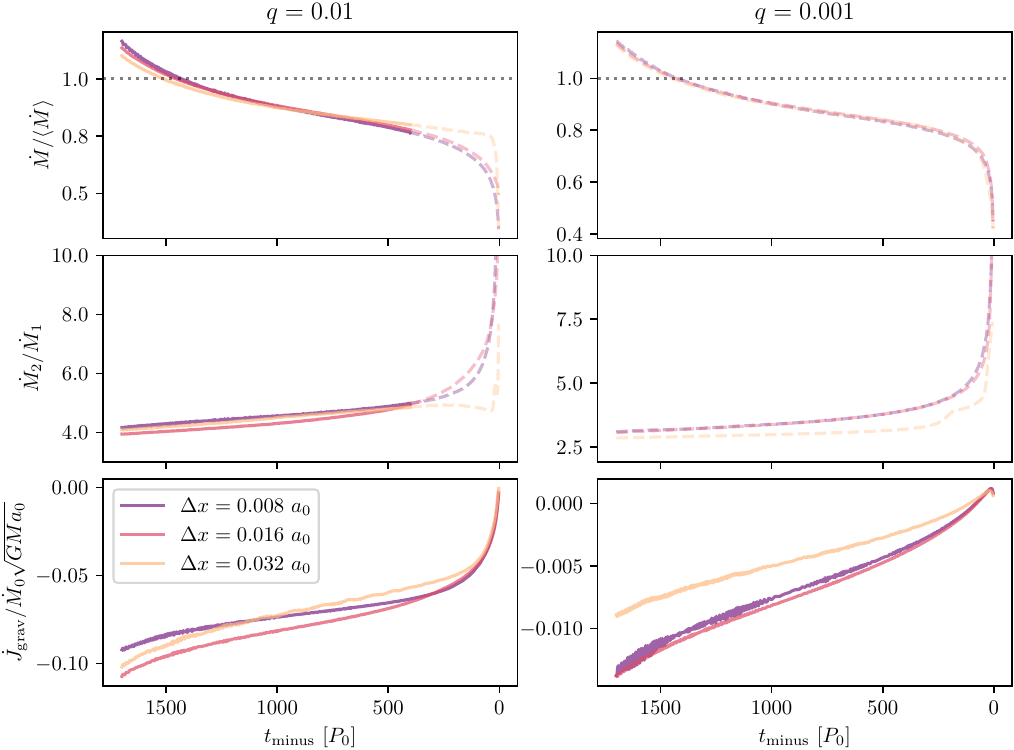}
    \caption{Time series of the mass accretion rates $\dot{M}$ (top row), the preferential accretion rate $\dot{M}_2 / \dot{M}1$ (middle row), and the total gravitational torque $\dot{J}{\rm grav}$ (bottom row) for distinct runs with grid spacings $\Delta{x}$ in the range of $0.008a_0 - 0.032a_0$. We show numerical convergence for two mass ratios, $q = 0.01$ (left column) and $q = 0.001$ (right column), with the orbital Mach number set to $\mathcal{M} = 10$ for both studies. The accretion rate curves start as solid lines and transition to dashed lines when the sink radius of the secondary BH becomes comparable to the size of the Hill radius, indicating that gas flows in the secondary disk are not well resolved for the dashed curves. The accretion rates are normalized by the empirical average binary accretion rate throughout the inspiral, $\langle \dot{M} \rangle$, and the dotted line in the top row indicates where $\dot{M} = \langle \dot{M} \rangle$.}
    \label{fig:Mdot-Jdot-diff_res}
\end{figure*}

\subsection{Tidal Squeezing of the Inner Disk}
\label{subsec:jdot}
As discussed in Sec. \ref{subsec:snowplow}, the theory of disk-satellite interactions implies that the orbiting secondary generally exerts a negative gravitational torque to the inner disk. When prescriptions for that torque are utilized in a one-dimensional disk model, they can lead to the formation of a zero-density, insulating gap in the disk surrounding the secondary. If the secondary is on a GW-driven inspiral, the insulating gap (which prohibits the flow of gas across the secondary's orbit) then implies tidal squeezing of the inner disk (so-called ``snowplow'' mechanism), and a potentially dramatic enhancement of the accretion rate $\dot M_1$ to the primary in the very late inspiral. In Sec. \ref{subsec:snowplow}, we referenced other studies that have observed that the insulating gap does not form in two-dimensional hydrodynamics calculations, and we also pointed out that the magnitude of the gravitational torque denoted as $\dot J_{\rm plow}$, needed to reduce the inner disk radius at the rate $\dot a_{\rm gw}$, diverges in the late inspiral. We report here the time series of gravitational torque $\dot J_{\rm in}$, applied to the binary by the inner disk (as well as $\dot J_{\rm tot}$, the torque from the whole disk) and confirm our estimate from Sec. \ref{subsec:snowplow}, that in the final thousands of orbits, $\dot J_{\rm in}$ is exceeded by $\dot J_{\rm plow}$, meaning that tidal squeezing becomes ineffective well before the merger.

Fig. \ref{fig:jdotin} shows time series data from the gravitational torques on the binary for three mass ratios ($q = 0.002, 0.01, 0.02$). The bottom panel shows $\dot J_{\rm in}$ (solid) and $\dot J_{\rm plow}$ (dashed) as derived in Eqn. \ref{eq:Jdotplow}. The filled in circles are where $\dot{J}_{\rm in} = \dot{J}_{\rm plow}$. Notice that $\dot J_{\rm plow}$ diverges close to merger; the secondary black hole can not provide enough torque to squeeze the inner disk in the final $\sim\!4000$ orbits, consistent with the estimate given in Sec. \ref{subsec:snowplow}. The top panel of Fig. \ref{fig:jdotin} confirms an assumption we made in that estimate, that the non-dimensional torques do not change radically throughout the inspiral.

\subsection{Numerical Convergence}
\label{subsec:numerical}

We ran many simulations to check the sensitivity of our results to numerical parameters, including the grid resolution and the start time of the simulation relative to the binary merger; see Appendix \ref{sec:appendix} for additional discussions on the sensitivity of our results to the sink size and sink rate. Fig. \ref{fig:Mdot-diff_tstart} presents the accretion rates of the fiducial binary for four simulations with different start times. The startup transients settle after roughly a hundred to a thousand orbits, and henceforth the accretion rates show very little sensitivity to the simulation start time. Also, in Fig. \ref{fig:Mdot-diff_tstart}, we have plotted the total accretion rate for a fiducial binary that is not inspiraling. This also helps assure us that the long-term viscous decoupling discussed in Sec. \ref{subsec:mdots} is uniquely a consequence of the inspiraling binary.

We also checked that our accretion rate and gravitational torque time series data are well converged with respect to the grid resolution. In Fig. \ref{fig:Mdot-Jdot-diff_res}, we have plotted the total accretion rate, preferential accretion rate, and gravitational torque for simulations performed with different resolutions and for two different mass ratios. Resolution is quantified by the grid spacing $\Delta x / a_0$ (better resolution means smaller $\Delta x$). Some disagreement is seen in both $\dot M_1$ and $\dot M_2$ and the gravitational torques when the grid spacing is greater than $\Delta x \simeq 0.016a_0$, especially in the very late inspiral stage when the distance between the binary components becomes poorly resolved. However, at early times, the shapes of the accretion rate curves, as well as the torques, are reasonably consistent with one another when the grid spacing is smaller than $\Delta x \simeq 0.016 \ a_0$. Additionally, we see convergence for both the fiducial mass ratio $q=0.01$ and for the lowest mass ratio tested $q=0.001$ in both the accretion rates and torques in Fig. \ref{fig:Mdot-Jdot-diff_res}.

\section{Electromagnetic Signatures}
\label{sec:observables}
The inspiral of unequal mass MBHB\ap{s} might produce tell-tale EM signatures. The detections of such EM signatures by ground or space-based observatories 
especially alongside a GW detection by \textit{LISA}, would provide an abundance of knowledge on binary evolution, accretion, and environments of MBHB\ap{s} within galaxies. To maximize the likelihood of an EM detection, we must understand how the hydrodynamical interactions between the binary and disk affect the thermal radiation released from the accretion flow. In this section, we use the spectral toy model from Sec. \ref{subsec:spectrum}, combined with the measured mass accretion rates, to provide a reasonable first approximation of the brightness and color evolution of an accretion flow surrounding an inspiraling black hole binary of significantly unequal mass. Note that a full model of the variable AGN emission around the time of a merger will require a more detailed treatment of the radiative transfer, both within the disk and also throughout the larger environment. Nevertheless, we do believe the analysis presented here qualifies as a first approximation of the variable AGN disk emission before and after a MBHB merger.

\subsection{Spectral Evolution and Light Curves}
\label{subsec:spectralevolution}
\begin{figure}
    \centering
    \includegraphics{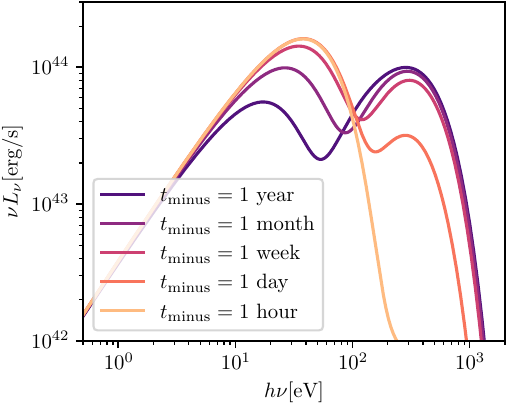}
    \caption{Evolution of the total spectral energy distribution of thermal emission produced by the fiducial system from $1$ year before merger (purple) to $1$ hour before merger (yellow). The model assumes that the accretion flow can be characterized as a steady state $\alpha$-disk, with a radial temperature profile given by Eqn. \ref{eq:temp}.}
    \label{fig:spectrum_evolution}
\end{figure}

In Fig. \ref{fig:spectrum_evolution} we plot the composite spectral energy distribution of the accretion flow in the fiducial binary simulation, at different times throughout the inspiral. We used the three-disk toy model from Sec. \ref{subsec:spectrum}, in which the spectrum is computed by approximating each disk (inner, secondary, and outer) to emit as a blackbody. The inner cutoff radii for the inner and secondary disk are $R_i = 6GM_i / c^2$, the innermost stable circular orbits of the respective black hole components. The inner radius of the outer disk and the outer radius of the inner disk are set to the instantaneous separation of the binary, while the secondary disk is extended to the instantaneous Hill radius. The result is not sensitive to the outer radius of the outer disk.

In Fig. \ref{fig:LC}, we plot the soft UV (top), hard UV (second), soft X-ray (third) and hard X-ray (bottom) light curves derived from our three-disk toy model of the disk temperatures. We define the soft UV energy range as $1.5-4$ eV, the hard UV energy range as $4-100$ eV, the soft X-ray energy range as $0.2-2$ keV, and the hard X-ray energy range as $2-12$ keV. 
Note that once the binary reaches the decoupling separation, we have manually imposed in our model to evolve the outer disk cavity wall at the viscous drift rate rather than at the binary merger rate to highlight the difference in the EM signatures pre- and post-viscous decoupling.
This is the reason for the abrupt change in the light curves seen around a couple of days prior to the merger, at the vertical dashed line corresponding to the decoupling timescale in Fig. \ref{fig:LC}.

Years before merger, the majority of the overall accretion power is released as X-rays from the small secondary disk. However, as the inspiral proceeds, the outer disk reaches deeper into the gravitational potential of the primary, glowing brighter in the UV, while the secondary is gradually starved and gets dimmer in the X-rays. The outer disk follows the inspiraling secondary black hole, and as it does so, the inner edge of the outer disk increases in speed and heats up the gas, leading to a brightening effect. The UV brightening peaks at the time when the disk viscously decouples from the binary as seen in Fig. \ref{fig:spectrum_evolution} at $t_{\rm minus} \simeq 1$ day. For the fiducial disk-binary model, the X-ray emission is surpassed by the UV emission on the order of one month before the merger. At the same time as the peak of the UV emission, the X-ray emission from the secondary disk decays because of the fast reduction in the rate of mass supplied to the secondary very late in the inspiral \citep{Krauth2023b}. This is also evident in the measured accretion rates in Fig. \ref{fig:Mdot}.

We do not show the post-merger EM signatures associated with the remnant fueling curves presented in Sec. \ref{subsec:remnant}, because the disk is far from being viscously relaxed while the accretion rate recovers to $\dot M_0$. Analytic estimates for this phase are presented in \cite{Milosavljevic2005, Schnittman2008, Shapiro2010}. We also do not show the orbital period variations in the light curves or spectrum, as these cannot generally be captured in the three-disk model. Spectral evolution on the time scale of individual binary orbits requires a self-consistent thermodynamic treatment and will be presented in future work (see Sec. \ref{subsec:future}).

\subsection{Prospects for EM Detections}
\label{subsec:telescopes}
Here we consider how our results might influence proposed strategies to discover the EM counterparts of \textit{LISA} GW detections. As discussed in the introduction, EM counterparts from \textit{LISA} sources will likely be needed to confidently identify the host galaxies of MBHB inspirals.

Our results indicate that in the years before a merger, accreting unequal-mass black hole binaries get brighter in the UV and dimmer in the X-ray. They also exhibit an abrupt disappearance in X-rays in the final hours to days before the merger. We further predict that variations at the orbital period should be discernible at the tens-of-percent level, resulting from orbital-period modulations of the accretion onto the secondary. Observatories carrying out time domain surveys, or space observatories capable of slewing in response to a GW trigger, could thus be instrumental in EM counterpart identification. 
We discuss below the relevance of these three predictions (periodicity, UV brightening, X-ray dimming) to current and upcoming observatories.

With regard to periodicity in the EM signatures,
the \textit{Nancy Grace Roman Space Telescope}, a wide-field optical to near-infrared space telescope, could potentially detect these EM signatures in an accreting MBHB, corresponding to the quasi-periodic variability of the mass accretion rates shown in the insets of Figs. \ref{fig:Mdot2insetq0.01} and \ref{fig:Mdot2insetq0.1}. Such periodicity detections by \textit{Roman} could become extremely valuable if the system is $10-20$ years away from merging because the merger might then take place during the \textit{LISA} period of operation \citep{WhitePaper2023}. As discussed in Sec. \ref{subsec:fiducial}, a $10^7$ M$_\odot$ binary that is 50 years away from merging has a rest-frame orbital period of $\unit[1.8]{days}$ and in Fig. \ref{fig:Mdot2insetq0.01}, the period is seen to decrease to $\unit[0.9]{days}$, around $10$ years before merging. At $z \sim 1-2$, this corresponds to multi-day periodicity, around a factor of $2$ shorter than the anticipated $\sim\!5$ day cadence planned for the \textit{Roman} high latitude time domain survey. Our calculations in Sec. \ref{subsec:variability} only predict the periodic variations of the mass accretion rates, so follow-up calculations will be needed to predict the level of periodicity expected where \textit{Roman} is sensitive. This is an intended area of future work; see Sec. \ref{subsec:future}.

\begin{figure}
    \centering \includegraphics{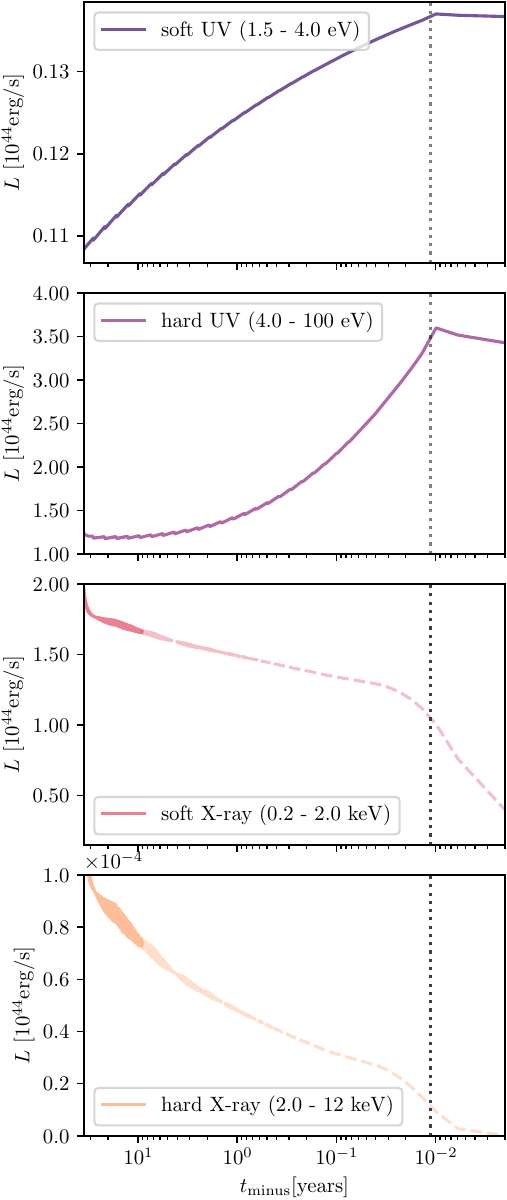}
    \caption{Light curves of the soft UV ($\unit[1.5-4]{eV}$), hard UV ($\unit[4-100]{eV}$), soft X-ray ($\unit[0.2-2]{keV}$), and hard X-ray ($\unit[2-12]{keV}$) emission, obtained by integrating the luminosity found in Fig. \ref{fig:binary_spectrum} over each energy band at each time. Light curves are shown for the fiducial binary where the mass ratio is $q = 0.01$ and the orbital Mach number is $\mathcal{M} = 10$. The X-ray light curves start as solid lines and transition to dashed lines when the sink radius of the secondary BH becomes comparable to the size of the Hill radius, indicating that gas flows in the secondary disk are not well resolved for the dashed curves. The dotted vertical line corresponds to the viscous decoupling timescale in Eqn. \ref{eq:t_dec}.}
    \label{fig:LC}
\end{figure}

The primary EM signature predicted in this work is a gradual increase in UV brightness from the circumbinary disk, aligning with the contraction rate of the binary driven by gravitational radiation emission.
The Legacy Survey of Space and Time (\textit{LSST}) of the \textit{Vera Rubin} Observatory
will observe nearly half the sky, $\unit[18,000]{deg^2}$, and will detect up to $10^6$ galaxies per square degree, observing quasars with a cadence of once per day \citep{Xin2021, Xin2024}. \textit{Vera Rubin's} \textit{LSST} camera will be sensitive in the soft UV energy range up to ~$4$ eV. In the top panel of Fig. \ref{fig:LC}, we have plotted the light curve for the approximate \textit{LSST} sensitivity. The steady UV brightening we are predicting to accompany a \textit{LISA} inspiral could be evident in \textit{LSST} archival data over the $10$ years prior to a GW detection by \textit{LISA}; however, the UV-brightening is more pronounced and thus easier to detect in the hard UV energy ranges (second panel of Fig. \ref{fig:LC}). Therefore, telescopes in the hard-UV energy range, such as the recently selected NASA Explorer mission, \textit{UVEX}, could be better suited to detect the predicted UV brightening \citep{Kulkarni2021}. Alternatively, since the characteristic temperature of the disk's inner edge decreases with the black hole mass, it is possible that \textit{LSST} will be sensitive to the variable emission from higher-mass AGNs.


The X-ray dimming after viscous decoupling predicted in equal mass binaries by \cite{Krauth2023b} and confirmed for unequal mass binaries here could be detectable by the \textit{Athena} X-ray telescope and/or \textit{eROSITA}. \textit{Athena} could potentially operate during the lifetime of \textit{LISA} to have simultaneous multimessenger detections, while \textit{eROSITA} is a current X-ray detector that could provide a reference sky template to aid in host galaxy characterization \citep{Merloni2012}. In the bottom two panels of Fig. \ref{fig:LC}, we have plotted the light curves for both the soft and hard X-ray energy ranges. The soft X-rays will be brighter initially, making them easier to observe; however, both energy ranges exhibit a significant drop in brightness due to viscous decoupling.

\section{Summary}
\label{sec:discussion}
\subsection{Main Results}
\label{subsec:results}

Using analytic estimates and high-resolution grid-based hydrodynamics simulations, we have explored the gas dynamics and EM signatures of accreting binary black holes with significantly unequal masses, $q = 10^{-3} - 10^{-1}$ and $M \sim 10^7 M_\odot$, over decade-like timescales before a merger. Our aim is to help enable EM counterpart detection and host galaxy identification of binary black hole inspirals observed by \emph{LISA}. Our main results are summarized here:
\begin{enumerate}
    \item
    The viscous decoupling time is generally on the order of $\tau_{\rm dec}\sim$ days for $\sim 10^7 M_\odot$ binaries. One might reasonably expect the binary accretion rate $\dot M$ to remain steady until $\sim \tau_{\rm dec}$ before merger; however, our simulations show significant reductions in $\dot M$ over decade-like time scales, $10^3 \tau_{\rm dec}$. We attribute this effect to the decrease of angular momentum current through the disk, accompanying the binary contraction. A model of this process is the subject of a forthcoming work.
    \item
    Low mass ratio binaries show accretion rate variability at the orbital period, at the level of around $10\%$. 
    Stochastic variability is reduced throughout the inspiral when the mass ratio is $q=0.01$ compared to $q=0.1$, consistent with the trend reported in \cite{Dittmann2024}.
    The secondary generally consumes most of the mass coming to the binary (preferential accretion effect). Both components are gradually starved of mass supply as the inspiral progresses, but the secondary starves more slowly than the primary.
    \item
    EM flares produced by tidal squeezing of the inner disk (snowplow effect) are unlikely. This is based on dynamical considerations and confirmed in the simulations; the gravitational torque becomes too weak years before merger to reduce the orbits of inner disk gas parcels faster than $\dot a_{\rm gw}$. Instead the inner disk gas accretes to the secondary during the inspiral, leading to the observed slower starvation of the secondary.
    \item
    We developed a toy model for the quasi-thermal emission spectrum of the accretion flow surrounding a low mass ratio binary black hole based on the assumption that the outer, inner, and secondary disks are viscously relaxed. The simple model predicts an emission spectrum with peaks in the UV and X-ray, reflecting the characteristic temperatures of the outer and circum-secondary disks, respectively. Those disks out-shine the inner (primary) disk because of the preferential accretion effect.
    \item
    Together with the mass accretion rates measured in the simulations, the toy model predicts a decade-like gradual brightening in the UV, and dimming in the X-ray, followed by an abrupt disappearance of the X-rays around time $\tau_{\rm dec} \sim \unit[]{days}$ before the merger \citep[analogous to][]{Krauth2023b}, and finally year-like gradual X-ray re-brightening associated with fueling of the remnant black hole. The long-term pre-merger UV brightening is the result of the increasing peak temperature of the outer disk, as it extends deeper into the potential well of the primary throughout the inspiral.
\end{enumerate}

\subsection{Caveats and Future Work}
\label{subsec:future}
Our simulations are based on two-dimensional (vertically averaged) solutions of the Navier-Stokes equations in the local isothermal approximation; the effects of shock heating, radiation, magnetic fields, general relativity, and out-of-plane gas flows, including winds and jets, are all neglected. Other groups are actively exploring these, especially in the context of near-equal mass ratio binaries in the final tens of orbits before a merger (see references given in the introduction). The simplified simulations presented here were necessary to study the long-term dynamics and EM signatures and may be accurate enough to help identify host galaxies of \emph{LISA} events.

In future work, we will present simulations of unequal-mass inspirals with a more realistic thermodynamics treatment to properly account for shock heating, as done with the \texttt{Sailfish} code in \cite{WesternacherSchneider2022, WesternacherSchneider2023, Krauth2023a, Krauth2023b, DeLaurentiis2024}. In the context of near-equal mass binaries, shock-heating does significantly alter the profile of disk surface temperature \citep[e.g.][]{Farris2015}. However, when the mass ratio is small, we expect less of the gas orbital energy to be dissipated in strong shocks, so the result might not be very different from the spectrum toy model we presented here.

We do not expect general relativistic effects to significantly alter the long-term brightness and color changes of a binary-host quasar in the years before a merger because the long-term changes are driven by large-scale viscous relaxation of the disk far from the gravitational radii of either black hole. 
Accounting for magnetic fields and radiation could change some of our predictions, especially in a regime where the mass flow to the binary is comparable to the Eddington rate of the primary. In such cases, preferential accretion implies the secondary tries to accrete much faster than its own Eddington rate and that radiation-driven outflows could significantly change the global mass currents in the system.

\section*{Acknowledgments}
M. Clyburn acknowledges support from the NASA Future Investigators Program (FINESST) through Award No. 80-NSSC-23K1443.
J. Zrake acknowledges support from the \emph{LISA} Preparatory Science Program (LPS) through NASA Award No. 80-NSSC-24K0440.
J. Zrake also acknowledges valuable discussions with Alex Dittmann, Sterl Phinney, Julian Krolik, Elena Rossi, and Zoltan Haiman, some of which took place at the 2022 binary accretion workshop hosted by the Kavli Institute for Theoretical Physics (KITP);
this research was supported in part by grant NSF PHY-1748958 to KITP.
The authors would also like to thank the anonymous referee for their detailed reading of our manuscript and constructive input.
All simulations were performed on Clemson University's Palmetto cluster.

\section*{Data Availability}

The data underlying this article will be shared on reasonable
request to the corresponding author.



\bibliographystyle{mnras}
\bibliography{main}




\appendix

\section{Sensitivity to Numerical Parameters}
\label{sec:appendix}

\begin{figure}
    \centering
    \includegraphics{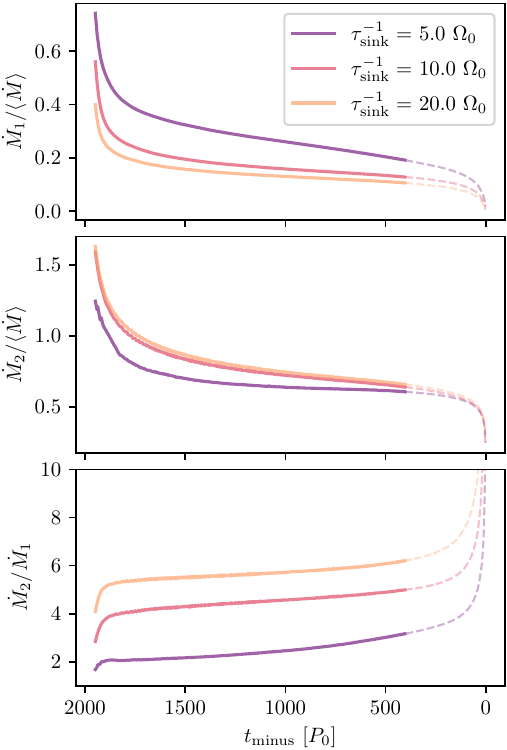}
    \caption{Time series of the mass accretion rates $\dot{M}_1$ (top), $\dot{M}_2$ (middle), and $\dot{M}_2/\dot{M}_1$ (bottom) for a binary mass ratio of $q = 0.01$ and a gas orbital Mach number of $\mathcal{M} = 10$, with varying sink rates and a sink size of $r_{\rm sink} = 0.05 a_0$. Each line starts as a solid curve and transitions to a dashed curve at the time when the secondary BH sink radius becomes comparable to the size of the Hill radius, beyond which the gas flow in the secondary disk is not well resolved for the dashed curves.}
    \label{fig:Mdot-sink-rate}
\end{figure}

\begin{figure}
    \centering
    \includegraphics{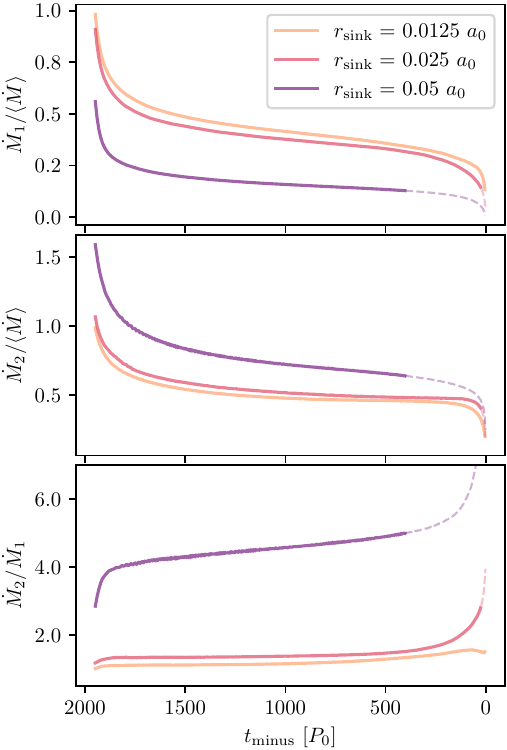}
    \caption{Time series of the mass accretion rates $\dot{M}_1$ (top), $\dot{M}_2$ (middle), and $\dot{M}_2/\dot{M}_1$ (bottom) for a binary mass ratio of $q = 0.01$ and a gas orbital Mach number of $\mathcal{M} = 10$, with varying sink sizes and a sink rate of $\tau_{\rm sink}^{-1} = 10 \ \Omega_0$. Each line starts as a solid curve and transitions to a dashed curve at the time when the secondary BH sink radius becomes comparable to the size of the Hill radius, at which point the gas flow in the secondary disk is not well resolved for the dashed curves.}
    \label{fig:Mdot-sink-size}
\end{figure}

As mentioned in Sec. \ref{subsec:Equations}, the accretion rates of low mass ratio binaries are sensitive to the sink size of the black holes and should be interpreted with caution. In Figs. \ref{fig:Mdot-sink-rate} and \ref{fig:Mdot-sink-size}, we plot the primary, secondary, and preferential accretion rates for different sink rates and sink radii, respectively. All simulations discussed in this section use a binary mass ratio of $q=0.01$ and an orbital Mach number of $\mathcal{M}=10$.

In Fig. \ref{fig:Mdot-sink-rate}, we present simulations with sink rates of $\tau_{\rm sink}^{-1} = 5, 10, 20 \ \Omega_0$, each using a sink size of $r_{\rm sink} = 0.05 \ a_0$. For the results presented in Sec. \ref{sec:results}, we use a sink rate of $\tau_{\rm sink}^{-1} = 10 \ \Omega_0$, which, according to Fig. \ref{fig:Mdot-sink-rate}, is high enough that the accretion rates are relatively insensitive to the sink rate. However, the accretion rate to the primary BH becomes spuriously large, and the accretion rate to the secondary BH becomes spuriously small when the sink rate is too low, i.e., $\tau_{\rm sink}^{-1} < 10 \ \Omega_0$. Note that the preferential accretion rate increases with increasing sink rate. Although the components' accretion rates are nearly identical at high sink rates, the preferential accretion depends on the ratio $\dot{M}_2 / \dot{M}_1$. Consequently, even small variations in $\dot{M}_1$ between runs with different sink rates can lead to appreciable changes in the preferential accretion.

In Fig. \ref{fig:Mdot-sink-size}, we present simulations with sink radii of $r_{\rm sink} = 0.05, 0.025, 0.0125 \ a_0$, each with a sink rate of $\tau_{\rm sink}^{-1} = 10 \ \Omega_0$. As mentioned in Sec. \ref{subsec:Equations}, all simulations in this manuscript begin with $r_{\rm sink} < r_{\rm hill}$, but during inspiral, the Hill radius decreases, making the flow of gas around the secondary increasingly difficult to capture. For low mass ratios, the accretion rate to the primary BH becomes spuriously small, and the accretion rate to the secondary BH becomes spuriously large when the sink size is too large. Thus, the preferential accretion in the bottom panel of Fig. \ref{fig:Mdot-sink-size} is enhanced for smaller BH sinks. Nevertheless, the sensitivity of the accretion rates to the sink size does not affect our key results of a long-term viscous decoupling and a gradual UV brightening during inspiral.



\bsp	
\label{lastpage}
\end{document}